\documentclass[11pt]{article}
\usepackage{amsmath,sint,epsfig,macros,cite}
\usepackage{macros_static}
\usepackage{textcomp}

\begin{document}

\begin{titlepage}

\begin{flushright}
\vskip 0.7cm
DESY 05-082\\
HU-EP-05/22\\
SFB/CPP-05-19
\end{flushright}

\vskip 0.95cm
\begin{center}
{\Large\bf 
On lattice actions for static quarks
\\[0.5ex] 
}
\end{center}
\vskip 0.35cm
\vbox{
\centerline{
\epsfxsize=2.8 true cm
\epsfbox{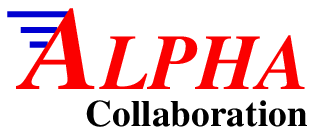}}
}
\vskip 0.1cm
\begin{center}
{
Michele Della Morte$^{\scriptscriptstyle a}$,
Andrea Shindler$^{\scriptscriptstyle b}$ and  
Rainer Sommer$^{\scriptscriptstyle c}$
}
\vskip 0.5cm
{
$^{\scriptstyle a}$ 
Institut f\"ur Physik, Humboldt Universit\"at, \\
Newtonstr. 15, 12489 Berlin, Germany 
\vskip 2.0ex
$^{\scriptstyle b}$ NIC/DESY,\\
Platanenallee 6, 15738 Zeuthen,  Germany
\vskip 2.0ex
$^{\scriptstyle c}$
DESY, \\
Platanenallee 6, 15738 Zeuthen, Germany
\vskip 2.0ex
}
\vskip 0.775cm
{\bf Abstract}
\vskip 0.1ex
\end{center}
We introduce new discretizations of the action for static quarks.
They achieve an exponential improvement (compared to the Eichten-Hill 
regularization) on the signal to noise ratio
in static--light correlation functions. This is explicitly checked in a
quenched simulation and it is understood quantitatively
in terms of the self energy of a static quark and the lattice
heavy quark potential at zero distance. 
We perform a set of scaling
tests in the Schr\"odinger functional and find scaling violations in the 
O($a$) improved theory to be rather small -- for one observable 
significantly smaller than with the Eichten-Hill regularization.
In addition we compute the improvement coefficients of the static light axial
current up to $\rmO(g_0^4)$ corrections
and the corresponding renormalization
constants non-perturbatively. The regularization dependent part of the 
renormalization of the b-quark mass in static approximation is also 
determined.

\vskip 0.29cm
\vfill

\begin{center}
June 2005
\end{center}

\eject
\vfill
\eject

\end{titlepage}

\newcommand{\ssect}[1]{\noindent {\bf  #1.}}

\section{Introduction}

Lattice QCD is playing an important r\^ole in the interpretation
of B-physics experiments~\cite{CKM:CERN}. However, as 
new physics is hiding behind the standard 
model, the required precision for lattice computations of B-meson
transition amplitudes is becoming more and more demanding.
The development of new algorithms~\cite{algo:GHMC,algo:GHMC3,algo:L1,algo:L2}
promises that light dynamical fermions at small lattice
spacings, $a$, can be reached in the near future, which will reduce one
of the most important systematic errors.

Still, the treatment of b-quarks on the lattice remains another
difficult part of these computations, since it appears unlikely 
that lattice spacings small enough to satisfy  $a<1/\mbeauty$ will soon come into
reach. We refer the reader to reviews for an explanation
of the various approaches to this 
problem~\cite{lat91:lepage,lat03:kronfeld,reviews:physcoll}.

A theoretically clean solution is provided by HQET. 
This effective theory starts from the static approximation 
describing the asymptotics as $\mbeauty \to \infty$. 
Corrections $\rmO(1/\mbeauty)$ have to be computed by a $1/\mbeauty$
expansion, where the higher dimensional interaction terms 
in the effective Lagrangian are treated  {\em as insertions} into static correlation functions. 
An attractive feature of this approach is that the continuum limit
exists and results are independent of the regularization.
The theory requires non-perturbative renormalization, but a concrete 
way to carry this out has been proposed and tested for a 
simple case \cite{hqet:pap1}. 

Furthermore also just the leading order term (static)  is of considerable interest,
since it provides a limit of the theory, which in many cases is not
expected to be far from results at the physical point. Other methods
to treat heavy quarks can thus be tested by checking whether they 
smoothly approach the static limit $1/\mbeauty \to 0$. In fact, results can
also be obtained from interpolations in $1/\mbeauty$ 
between points below the b-quark mass
and the static limit, see \cite{lat03:juri} for a precise demonstration. 

As we will discuss in detail in \sect{s:noise}, the noise-to-signal ratio
of static-light correlation functions grows exponentially in Euclidean time, $\sim \exp(\mu x_0)$
and the rate $\mu\sim1/a$ diverges as one approaches the continuum limit.
Good statistical precision is difficult to 
reach for $x_0>1\,\fm$ \cite{stat:eichten,stat:hashi}.
In the past, sophisticated wavefunction techniques have been
used to obtain ground state properties 
\cite{fbstat:old1,fbstat:old2,stat:fnal2}.
The results of these attempts, obtained with the standard Eichten-Hill
regularization~\cite{stat:eichhill1}, were not completely satisfactory 
in all cases~\cite{reviews:beauty}.
In \cite{stat:letter} we used the fact that $\mu$ is not universal. By changing the regularization,
we were able to obtain results at large $x_0$, where the ground state can be isolated with good
statistical and systematic precision. Here we will discuss these alternative actions
in more details. Particular emphasis is put on the size of discretization errors,
since their smallness should always be the first criterion for choosing an action.
Let us mention right away that we will consider only actions, which share the 
symmetries, eqs.~(\ref{e_spin},\ref{e_quarknumber}),
with the Eichten-Hill action, since these guarantee
that no $\Oa$ terms are needed in the static action to have $\Oa$ improvement.  
For all actions considered,
we determine the coefficients of the improvement terms needed for the 
axial current 
with 1-loop precision. 
In particular we correct for a mistake in the perturbative computation 
of the improvement coefficients
in \cite{stat:letter}, see the Erratum. Furthermore we determine the 
action-dependent 
parts of the renormalization of the static axial current and the b-quark mass.


Before coming to the description of the static actions, we list some
preliminaries.
As a probe of the theory, we will consider correlation functions 
defined by the Schr\"odinger functional (SF) 
\cite{SF:LNWW,SF:stefan1}. For the introduction of static quarks in the SF 
as well as any unexplained notation
we refer to \cite{zastat:pap1}.
We adopted the $\rmO(a)$-improved Wilson regularization~\cite{impr:SW,impr:pap1} 
with non-perturbatively
determined value of $c_{\rm SW}$ \cite{impr:pap3} for the 
light quarks. The gauge sector is 
discretized through the Wilson plaquette action. 

Below we will report also on results of Monte Carlo computations.
Our simulation parameters are summarized in \app{s:param}. 
All the results have been obtained in the quenched approximation,
a part of them  already appeared in \cite{stat:letter} and \cite{lat03:michele}.

\section{Static actions\label{s:act}}

Static quarks are introduced on the lattice through fermionic fields $\psi_{\rm h}$ 
and $\psibar_{\rm h}$,
which live on the lattice sites and satisfy the projection properties
\be
P_+ \heavy = \heavy \; , \quad\quad \heavyb P_+=\heavyb \;,  
\quad\quad P_+={{1+\gamma_0}\over{2}} \;.
\label{genaction}
\ee
The action $S_{\rm h}$ for static quarks has been derived by Eichten and 
Hill in~\cite{stat:eichhill1}. 
Here we adopt a notation, which allows us to provide a 
generalization of that action and write it in the form
\be
\label{Saction}
S_{\rm h}^{\rm W} =a^4 {{1}\over{1+a\; \delta m_{\rm W} }} \sum_x 
\heavyb(x) (D_0^{\rm W} + \delta m_{\rm W} ) \heavy(x) \;, 
\ee
with the covariant derivative
\be
D_0^{\rm W} \heavy(x) = {{1}\over{a}} \left[ \heavy(x) -W^{\dagger} (x-a\hat{0},0)
 \heavy (x-a\hat{0}) \right] \; ,
\ee
where $W(x,0)$ is a gauge parallel transporter with the gauge transformation
properties of the link $U(x,0)$. 
The Eichten-Hill action is given by setting $W(x,0) = U(x,0)$.
The quark propagator $ G_{\rm h}^{\rm W}(x,y)$  satisfying 
$(D_0^{\rm W}+\delta m_{\rm W})G_{\rm h}^{\rm W}(x,y)=\delta(x-y) P_+$, reads
\bes
\label{stat_proI}
\!\!\!\!\!\!\!\!\!\!\!\!\!G_{\rm h}^{\rm W}(x,y) &\!\!\!=\!\!\!& \theta(x_0-y_0) \;
 \delta({\bf{x-y}}) \;(1+a \; \delta m_{\rm W})^{-(x_0-y_0)/a}\; 
{\cal{P}}^{\rm W}(y,x)^{\dagger} \; P_+ \; , \\
\!\!\!\!\!\!\!\!\!\!\!\!\!{\cal{P}}^{\rm W}(x,x)&\!\!\!=\!\!\!&1 \; ,\\
\!\!\!\!\!\!\!\!\!\!\!\!\!{\cal{P}}^{\rm W}(x,x+R \hat{\mu}) &\!\!\!=\!\!\!& 
W(x,\mu)W(x+a \hat{\mu},\mu) \dots W(x+a(R-1) \hat{\mu},\mu) \;\; {\rm for \;\; 
R >0} ,
\label{stat_prop}
\ees
where $\delta m_{\rm W}$ cancels 
the divergence in the self-energy of the static quark (see below).
\subsection{Noise to signal ratio \label{s:noise}}
\begin{figure}[htb]
\hspace{0cm}
\centering
\epsfig{file=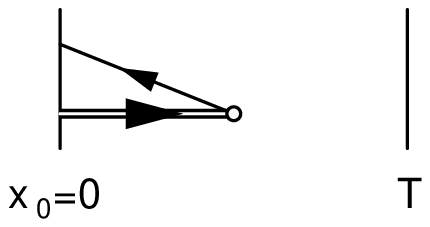,width=5.2cm,angle=0}
\epsfig{file=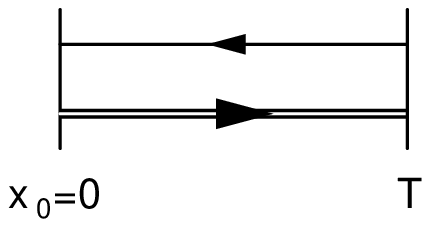,width=5.2cm,angle=0}
\vspace{-0.35cm}
\caption{\footnotesize{\sl Schematic representation of the correlation functions
$f_{\rm A}^{\rm stat}(x_0)$ and $f_{1}^{\rm stat}$ in the SF. The
double lines indicate the static quark propagators.}}
\label{fig_fA}
\end{figure}

In order to simplify
the following discussion we start from the action in \eq{Saction} 
but set $\delta m_{\rm W}=0$. This is possible since the 
dependence of correlation functions on $\delta m_{\rm W}$ 
is known exactly from the form of
the static quark propagator in eqs.~(\ref{stat_proI}-\ref{stat_prop})
and can be restored easily. The noise to signal ratio discussed here 
is independent of  $\delta m_{\rm W}$. As an example
we consider a correlation function of heavy-light fields such as 
$f_{\rm A}^{\rm stat}(x_0)$, which describes the
propagation of a static--light pseudoscalar meson.
Its definition 
\be
f_{\rm A}^{\rm stat}(x_0) = -{{a^6}\over{2}} \sum_{{\bf y},{\bf z}}
\langle A_0^{\rm stat}(x) \overline{\zeta}_{\rm h}({\bf y}) \gamma_5
\zeta_{\rm l}({\bf z}) \rangle \; ,
\label{fAstat}
\ee
involves the boundary fields $\overline{\zeta}_{\rm h}$ and $\zeta_{\rm l}$ 
(see \cite{zastat:pap1}) as well as the time component of the axial current
\be
A_0^{\rm stat}(x)=\psibar_{\rm l}(x) \gamma_0 \gamma_5 \psi_{\rm h}(x) \;,
\label{axial}
\ee
where the subscript `l' indicates the light quark fields.
The correlation function is represented pictorially in figure~\ref{fig_fA}.
%
From the quantum mechanical representation one expects the 
large $x_0$ asymptotic behavior
\be
f_{\rm A}^{\rm stat}(x_0) \propto e^{-E_{\rm stat} x_0} \;,
\label{QMrep}
\ee
where $E_{\rm stat}$ is the {\it binding energy} of the static--light system.
It depends on the choice for $W$ and diverges
approximately linearly in the lattice spacing. 
\be
E_{\rm stat} \sim \Eself + {\rm O}(a^0) \sim {{1}\over{a}} 
\eselfone \; g_0^2 + \dots \;.
\ee
This divergence is to be canceled
by $\delta m_{\rm W}=-\Eself + \rmO(a^0)$. The  finite part of $\delta m_{\rm W}$ 
is of course scheme dependent, but in perturbation theory one
can always perform the double expansion 
\bes
  \Eself &=& \Eself^{(1)}\, g_0^2 + \rmO(g_0^4) \, , \\
  \Eself^{(1)} &=& {{1}\over{a}} \eselfone +  \rmO(a^0) \,.
\ees
For  example, if we define $\Eself$ from a \SF correlation function
(see \eq{e:eselfsf}), the $a$-expansion appears as
an expansion in terms of $a/L$. The coefficient $\eselfone$ does
then depend on the action, but not on the correlation function
used to define it.

We now want to discuss the noise to signal ratio of the Monte Carlo
estimate of
$f_{\rm A}^{\rm stat}$. 
The variance of a generic observable $\cal O$ in a QCD simulation is given
by $\langle \tilde{\cal O}^2 \rangle_{\rm G}  - \langle \tilde{\cal O} 
    \rangle_{\rm G}^2$,
where the expectation value $\langle  ... \rangle_{\rm G}$ is
an average over the gauge fields including the weight obtained
after integrating out the fermion fields (the fermion determinant). 
$\tilde{\cal O}$ is derived
from $\cal O$ by performing the corresponding Wick contractions on each gauge
field background. 
In the  Monte Carlo one takes advantage of translation
invariance on the r.h.s. of \eq{fAstat} and estimates 
$f_{\rm A}^{\rm stat} = \langle \cal O \rangle$ with 
${\cal O} = -{{a^9}\over{2L^3}} \sum_{{\bf y},{\bf z},\vecx}
A_0^{\rm stat}(x) \overline{\zeta}_{\rm h}({\bf y}) \gamma_5
\zeta_{\rm l}({\bf z})$. Its variance,
$\sigma_{\rm A}(x_0)$,
can be rewritten as 
\bes
\sigma_{\rm A}(x_0) &=& {{a^{18}}\over {4 L^6}}
\sum_{{\bf x, y,  z,   x', y', z'}}
\left\langle \, [A_0^{\rm stat}]_{\rm hl}(x_0,\vecx) \, \overline{\zeta}_{\rm h}({\bf y}) \gamma_5
\zeta_{\rm l}({\bf z})  \times \nonumber 
\right.\\ && \left. 
\qquad \left\{[A_0^{\rm stat}]_{\rm h'l'}(x_0,\vecx') \,\overline{\zeta}_{\rm h'}({\bf y'}) \gamma_5
\zeta_{\rm l'}({\bf z'})\right\}^\dagger \, \right\rangle  \; - \; [\fastat(x_0)]^2\,,
\label{sigmaAstat}
\ees
where $\rm h'$ and  $\rm l'$ denote copies of  $\rm h$ and $\rm l$ differing only
by their flavor, which are introduced to be able to write the variance 
in the form of a standard expectation value.
Saturating $\sigma_{\rm A}(x_0)$ with intermediate states
shows that its large $x_0$ asymptotics is  
\be
\sigma_{\rm A}(x_0) \propto \sum_{\bf x, x'}\,
  C(\vecx,\vecx') \,\rme^{-x_0 E_{\rm ll'}(\vecx,\vecx')}\,,
\ee
with $E_{\rm ll'}(\vecx,\vecx')$ the energy of a state with
a static quark-antiquark pair at positions $\vecx,\vecx'$ 
and a light quark-antiquark pair with flavors $\rm l,l'$.
At large $x_0$ the sum is dominated by
the term with the smallest energy $E$. This is expected to
be given by $\vecx=\vecx'$, where the color charges of the static
quark and anti-quark compensate. The light quark-antiquark pair
should then feel little of the static quarks and its lowest energy
be  given approximately by $m_\pi$. This argument suggests
\be
  {\rm Min}_{\vecx,\vecx'}\, E_{\rm ll'}(\vecx,\vecx') =
  V(0) + m_{\pi} \,.
\ee 
Here $V(0)$ is the lattice static quark potential at zero distance,
which for example is computable  via the large $x_0$ asymptotics
\bes
\langle \, \heavyb(x)\gamma_5 \psi_{\rm h'}(x) \, 
       \overline{\zeta}_{\rm h'}({\vecx}) \gamma_5 \zeta_{\rm h}({\vecx}) 
        \, \rangle  \propto \rme^{-x_0 V(0)} \,.
\ees
The noise 
to signal ratio $R_{\rm NS}$ for
 $f_{\rm A}^{\rm stat}(x_0)$ should then approach
\be
R_{\rm NS} \propto e^{[E_{\rm stat}-(m_{\pi}+V(0))/2]\,x_0} \;.
\label{Lepf}
\ee
For the Eichten-Hill action (or more generally 
whenever $W(x,0)$ is unitary), the lattice potential vanishes
at zero distance and one obtains the formula given 
earlier by Lepage~\cite{lat91:lepage}. 
We infer  from \eq{Lepf} that the linear divergence in 
$E_{\rm stat}$ is responsible
for the exponential growth of the error with the Euclidean time $x_0$,
which has been observed in many numerical investigations 
of correlation functions such as 
$f_{\rm A}^{\rm stat}$ \cite{fbstat:old1,fbstat:old2,stat:hashi,stat:fnal2}.
It is then clear that the problem becomes more severe as the continuum limit
is approached.
In particular, for the Eichten-Hill action the coefficient $\eselfone$
was found to be rather large \cite{stat:eichhill_za} 
rendering precise computations hopeless.
Finally we remark again that the form of the 
static propagator shows that $R_{\rm NS}$ is (exactly) independent
of 
$\delta m_{\rm W}$. In the final formula \eq{Lepf} this is realized
since $E_{\rm stat}$ and $V(0)/2$ are shifted by the same amount,
\bes
\left.aE_{\rm stat}\right|_{\delta m_{\rm W}} &=& 
\left.aE_{\rm stat}\right|_{\delta m_{\rm W}=0} + \ln(1+a\delta m_{\rm W})
\nonumber \\
\left.aV(0)\right|_{\delta m_{\rm W}} &=& 
\left.aV(0)\right|_{\delta m_{\rm W}=0} + 2\ln(1+a\delta m_{\rm W}).
\nonumber
\ees
\subsection{Statistically improved discretizations}
A possible way to improve on the problem discussed in the previous section 
is to employ actions $S_{\rm h}^{W}$ 
inspired by variance reduction methods. 
Such methods led for example to the introduction of the one-link integral 
(or multihit) \cite{PPR} in the pure gauge theory. This provides unbiased 
estimators for quantities such as Polyakov loop correlation functions, 
significantly reducing at the same 
time their variance. Here, a similar change will not lead to 
an unbiased estimator, but rather has to be considered a change of the 
static action.
Alternatively (or simultaneously) one can try  to enhance the signal  by
choosing the parallel transporter $W$ such that $E_{\rm stat} \sim \Eself$ 
is comparatively small 
(assuming $V(0)$ to be numerically less relevant). 
In addition to this, we want to preserve on the lattice the following symmetries
of the static theory
\bi
 \item[i)] Heavy quark spin symmetry:
        \bes
          \label{e_spin}
          \heavy\longrightarrow {\cal V}\heavy\,,
          \qquad \heavyb\longrightarrow\heavyb {\cal V}^{-1}\,,\quad
          {\rm with} \quad
          {\cal V}=\exp(-i\phi_i \epsilon_{ijk}\sigma_{jk})\,,
        \ees
 \item[ii)] Local conservation of heavy quark flavor number:
        \be
          \label{e_quarknumber}
           \heavy\longrightarrow \rme^{i\eta(\vecx)}\,\heavy,
          \qquad \heavyb\longrightarrow\heavyb \rme^{-i\eta(\vecx)}.
        \ee
\ei
Together with gauge invariance, parity and cubic symmetry, this  is
enough to guarantee that the universality class and the O($a$)
improvement are unchanged with respect to the Eichten-Hill action,
as has been discussed in~\cite{zastat:pap1}.
Furthermore we want to keep the action as local as possible.
We therefore exclude constructions of $W$ which involve fields 
at a distance two lattice spacings or more away from the link 
$U(x,0)$. Taking these considerations into account,
we propose the following regularized actions
\bea
\label{SAPE}
S_{\rm h}^{\rm A} &: \quad W^{\rm A}(x,0) =& V(x,0) \; ,  \\
\label{Ss}
S_{\rm h}^{\rm s} &: \quad W^{\rm s}(x,0)  =& V(x,0) \left[ {{g_0^2}\over{5}}+\left(
{{1}\over{3}} \tr V^{\dagger}(x,0)V(x,0) \right)^{1/2} \right]^{-1} \; , \\
\label{SHYP}
S_{\rm h}^{\rm HYP} &: \quad W^{\rm HYP}(x,0)  =& V_{\rm HYP}(x,0) \; ,  
\eea
where $V(x,0)$ is the average of the six staples around the link $U(x,0)$
\bea
V(x,0) &=& {{1}\over{6}} \sum_{j=1}^3 \left[  U(x,j)U(x+a\hat{j},0)
U^{\dagger}(x+a\hat{0},j) \right.\nonumber \\ 
&&\qquad\; \left. + U^{\dagger} (x-a\hat{j},j)U(x-a\hat{j},0)
 U(x+a\hat{0}-a\hat{j},j) \right] \; ,
\eea
and $V_{\rm HYP}$ is the HYP link \cite{hyp}. In the latter case three coefficients
$(\alpha_1,\alpha_2,\alpha_3)\; \equiv \vec{\alpha}$ 
need to be specified in order to define the
combination of differently smeared links in the construction of the HYP link.
In the following we will only discuss the choices $\vec{\alpha}= (0.75,0.6,0.3)$,
 motivated in \cite{hyp} and $\vec{\alpha}=(1.0,1.0,0.5)$ obtained 
by an approximate minimization of $R_{\rm NS}$. They  define 
$S_{\rm h}^{\rm HYP1}$ and $S_{\rm h}^{\rm HYP2}$ respectively.
The construction of HYP links involves projecting $3 \times 3$
complex matrices onto SU(3). As we expect deviations from SU(3) to be small,
we prefer approximating  the projection by an
analytic function defined by the steps
\begin{equation}
W \rightarrow {{W}\,/\,{\sqrt{\tr (WW^{\dagger}) / 3  }  }}  \; , 
\end{equation}
followed by 4 iterations of
\begin{equation}
W \rightarrow X \left( 1-\frac{\rm i}{3} \Im \left(\det X\right)\right) \; ,
\quad {\rm with} \quad X=W\left(\frac{3}{2}-\frac{1}{2} W^{\dagger}W\right) \;.
\end{equation}
To be precise, this function is to be taken 
as part of the definition of $V_{\rm HYP}$, when our numerical
results for improvement coefficients and renormalization factors
are used in subsequent computations.

For the actions $S_{\rm h}^{\rm HYP1}$ and $S_{\rm h}^{\rm HYP2}$ the
value of $\eselfone$ has been computed in 
\cite{HYP:pot}. 
In the cases $W \in\!\!\!\!\! /$ SU(3) it is more convenient to look
at $\Eself-V(0)/2$ not only because this is the relevant quantity for 
the noise to signal ratio, but also because at 1-loop order the deviations of the
links $W$ from unitarity cancel in the combination. We indicate by $r^{(1)}$
the 1-loop coefficient in the perturbative expansion 
\be \label{e:r1}
 \Eself-V(0)/2 \sim \left({{1}\over{a}} r^{(1)} + {\rm O}(a^0)\right) \; g_0^2+ \rmO(g_0^4) \;.
\ee
Of course $r^{(1)}=\eselfone$ for 
$S_{\rm h}^{\rm HYP}$ and $S_{\rm h}^{\rm  EH}$.
In \app{s:PT} we outline a computation of 
$r^{(1)}$ for $S_{\rm h}^{\rm A}$  and
explain why it is the same for $S_{\rm h}^{\rm s}$ (at one-loop order).
%
%
\begin{table}[htb]
\centering
\begin{tabular}{c|ccc}
\hline\\[-2.0ex]
 $S_{\rm h}^{\rm W}$ & $r^{(1)}$ & $aE_{\rm stat}$& $aV(0)$\\[.5
ex]
\hline\\[-2.ex]
$S_{\rm h}^{\rm EH}$   & 0.16845(2)    & 0.68(9)    & 0.0       \\[.7ex]
$S_{\rm h}^{\rm A}$    & 0.05737(4)    & 0.85(2)    & 0.671(2)  \\[.7ex]
$S_{\rm h}^{\rm s}$    & 0.05737(4)    & 0.76(2)    & 0.496(2)  \\[.7ex]
$S_{\rm h}^{\rm HYP1}$ & 0.04844(1)    & 0.44(2)    & 0.0       \\[.7ex]
$S_{\rm h}^{\rm HYP2}$ & 0.03523(1)    & 0.41(1)    & 0.00001(1)       \\[.7ex]
\hline 
\end{tabular}
\caption{\footnotesize{\sl
One loop coefficients $r^{(1)}$, defined in
\eq{e:r1} and
non-perturbative values for $aE_{\rm stat}$ and $aV(0)$ at $\beta=6$.
}}\label{deltam}
\end{table}
%
Some results for $r^{(1)}$ are collected in table~\ref{deltam}. 
From the table we see that
perturbation theory already suggests  $S_{\rm h}^{\rm HYP2}$ 
as the most favorable choice concerning signal enhancement.

The discretizations $S_{\rm h}^{\rm A}$ and $S_{\rm h}^{\rm s}$ are 
inspired by 
noise reduction methods, namely APE smearing and one-link integral respectively. 
APE smearing was introduced in~\cite{smear:ape} where it was shown to suppress 
fluctuations in gauge invariant quantities in the pure gauge theory.
As shown by $r^{(1)}$, this can also be interpreted as a 
reduction of the static self energy. 
Finally $W^{\rm s}$ is an approximation of the SU(3) one-link
integral. For completeness we describe briefly how we arrived at 
$S_{\rm h}^{\rm s}$ in \app{s:onelink}. The reader may, however, take it just 
as another ansatz satisfying our criteria explained above.

\begin{figure}[htb]
\hspace{0cm}
\centerline{
\epsfig{file=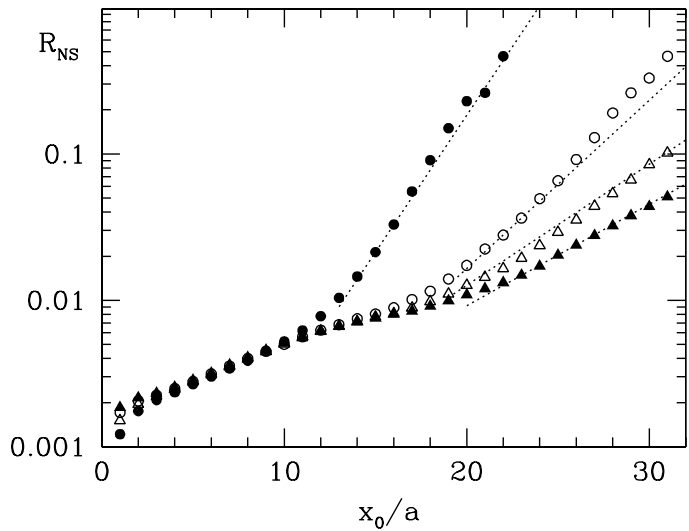,width=9.0cm,angle=0}
}
\caption{
\footnotesize{\sl
The ratio $R_{\rm NS}$ for the correlation function $f_{\rm A}^{\rm stat}(x_0)$
for a statistics of 5000 measurements on a  
$16^3 \times 32$ lattice at $\beta=6$.
Filled circles refer to $S_{\rm h}^{\rm EH}$, empty circles to $S_{\rm h}^{\rm A}$
 ($S_{\rm h}^{\rm s}$ gives similar results), 
empty (filled) triangles to $S_{\rm h}^{\rm HYP1}$ ($S_{\rm h}^{\rm HYP2}$).
}
\label{f_rns}}
\end{figure}

The effectiveness of these regularizations in reducing the noise to signal ratio
in static--light correlation functions has been checked non-perturbatively
by a simulation on a $16^3 \times 32$ lattice at $\beta=6/g_0^2=6$, 
where the lattice spacing is $a\approx\,0.1\,\fm$
(the hopping parameter
$\kappa$ has been set to the strange quark mass value, $\kappa=0.133929 $, 
as in~\cite{stat:letter}). 
Figure~\ref{f_rns} shows the results for $R_{\rm NS}$ 
obtained using an ensemble of almost 5000 configurations.

%
%
The largest
improvement is again given by the action $S_{\rm h}^{\rm HYP2}$, but we see
from the figure that all the proposed discretizations produce a clear signal
(for the considered statistics) at least up to a time separation of roughly
$2 \; {\rm fm}$.
The dotted lines in figure~\ref{f_rns} represent the predictions from 
\eq{Lepf}, where for $aE_{\rm stat}$, $aV(0)$ and 
$am_{\pi}$ we insert the estimates from the data. 
We summarize the obtained values of 
$aE_{\rm stat}$ and  $aV(0)$ in table~\ref{deltam}.
The formula in \eq{Lepf} turns out to be always quite 
accurate in describing the results. Furthermore, 
assuming $E_{\rm stat}-V(0)/2$ to be dominated
by its divergent term  and approximating
it by the leading perturbative estimate yields
the correct order for $R_{\rm NS}$.
It is thus to be expected 
that the ordering of $R_{\rm NS}$ observed in figure~\ref{f_rns}
will be preserved when going to smaller lattice spacings,
where it is even more important to keep $R_{\rm NS}$ at a reasonable
level if one wants to reach distances $x_0$ around $1-2$~fm. Indeed,
numerical experience supports this expectation \cite{stat:letter}.


\section{The size of discretization errors}
In addition to  the reduction of statistical errors,
the size of scaling violations has to be taken into account
when one chooses between different actions.
For this reason we designed a number
of scaling tests by which we could study the approach of a set of quantities
defined in the SF to their continuum limit values. 
Before going into the details of this study we first 
need to discuss the computation of the O($a$) improvement coefficients of the
 static axial  current for the different actions introduced.
\subsection{Computation of the improvement coefficients $c_{\rm A}^{\rm stat}$
and $b_{\rm A}^{\rm stat}$ }
The O($a$) improvement programme \cite{impr:Sym1,impr:Sym2} has been carried out for
 the EH action in~\cite{zastat:pap1}. In~\cite{stat:letter} the discussion has been
 extended to the actions we are considering here. As mentioned in the
 previous section, arguments based on the symmetries preserved
 on the lattice allow to
 show that the improvement pattern is the same in all cases.
In particular the actions are improved once the light sector has been
 improved; no new improvement terms are needed for the static part of the
 actions. 

Concerning the static--light axial density $A_0^{\rm stat} (x)$,
the improved version reads
\begin{equation}
 \label{e_aimpr}
 \Astatimpr \!=\! \Astat(x)\!+\!a\castat\delta\Astat(x)\;, \quad
  \delta\Astat(x) \!= \!\lightb(x)\gamma_j\gamma_5\!
  {\lnab{j}\!+\!\lnabstar{j} \over 2}\heavy(x)\;,
\end{equation}
where the improvement coefficient $\castat$ has been introduced.
In a mass independent renormalization scheme the renormalized density can be
written
\begin{equation}
(A_{\rm R}^{\rm stat})_0=\zastat(g_0,a\mu)(1+\bastat  a\mq)\Astatimpr\;,
\end{equation}
with $\mq$ the bare subtracted light quark mass (cf. \eq{qmass}),
 $\zastat(g_0,a\mu)$
the scale dependent renormalization factor of the axial
current and $\bastat$ a second improvement coefficient.
The values of $\castat$ and $\bastat$, as well as $\zastat$, depend on the choice for the
static action.
For the EH action the improvement coefficients $\castat$ and $\bastat$
have been determined at 1-loop order of perturbation theory 
in~\cite{zastat:pap1,MORNINGSTAR1,castat}, while the renormalization constant 
$\zastat$ has been computed
non-perturbatively in the SF scheme in~\cite{zastat:pap3}. 
In the remainder of this section we describe our computation of the
improvement coefficients $\castat$ and $\bastat$ for the actions in
 eqs.~(\ref{SAPE}~-~\ref{SHYP}), and we present a set of scaling studies.
We will come back to the renormalization constant $\zastat$ in section
4, where we also discuss the improvement the new actions can bring in
the computation of the b-quark mass following the strategy 
in~\cite{lat01:rainer,hqet:pap1}.
\subsubsection{Improvement conditions and results for $\castat$ and $\bastat$}
We want to compute the 1-loop coefficients  $c_{\rm A}^{\rm stat, (1)}$ 
and $b_{\rm A}^{\rm stat, (1)}$ of the improvement constants $\castat$ and $\bastat$.
To this end we have adopted a mixed strategy. For $S_{\rm h}^{\rm A}$ and 
$S_{\rm h}^{\rm s}$ the computation has been carried out 
analytically\footnote{At this 
order in perturbation theory the results for the improvement coefficients
are the same for the two regularizations, 
see appendix~\ref{s:PT}.}, 
while for $S_{\rm h}^{\rm HYP1}$ and $S_{\rm h}^{\rm HYP2}$
we have used Monte Carlo simulations to numerically estimate an effective 1-loop coefficient
for the coupling range relevant here. In both cases we have exploited the same improvement
conditions. We collect some details of the analytic computation in 
appendix~\ref{s:PT}.

\begin{figure}[htb]
\hspace{0cm}
\centering
\epsfig{file=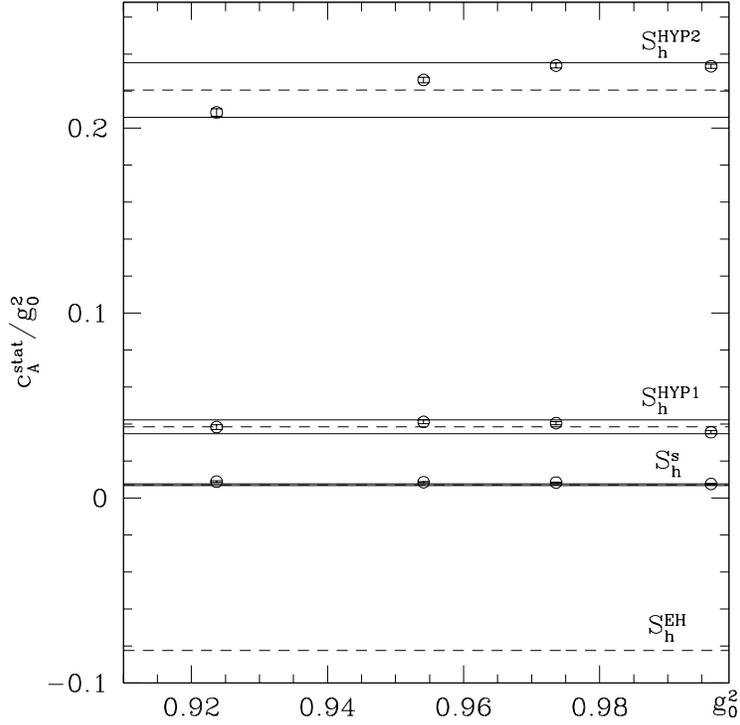,width=9.7cm,angle=0}
\caption{\footnotesize{\sl 
Numerical results for the improvement coefficient $\castat$. 
For $S_{\rm h}^{\rm EH}$ the result of \cite{castat} is plotted, 
for $S_{\rm h}^{\rm s}$
the band 
represents the result \eq{caA}, 
while otherwise it refers
to the numerical estimates eqs.~(\ref{caHYP1},~\ref{caHYP2}).
}}
\label{ca_fig}
\end{figure}

We introduce the correlation $\fastatimpr$ defined as $\fastat$ in~\eq{fAstat}
with the current $A_0^{\rm stat}$ replaced by the improved current in \eq{e_aimpr}.
Assuming the knowledge of $c_{\rm A}^{\rm stat}$
for a discretized action
$S_{\rm h}^1$, we have enforced the condition\footnote{Where necessary
  we explicitly indicate the dependence of the correlation functions
  and the improvement coefficients on the discretization of the static action.}
\begin{equation}
{\left.\fastatimpr(T/2,S_{\rm h}^1)\right|_{\theta}\over 
\left.\fastatimpr(T/2,S_{\rm h}^1)\right|_{\theta'}} \,=\,
{\left.\fastatimpr(T/2,S_{\rm h}^2)\right|_{\theta}\over 
\left.\fastatimpr(T/2,S_{\rm h}^2)\right|_{\theta'}}\;, 
\quad \theta=0,\; \theta'=1\;, \quad m_{\rm q}=0 \;,
\label{caimp}
\end{equation}
and solved the implicit relation for $c_{\rm A}^{\rm stat}$
of $S_{\rm h}^2$.
Expanding \eq{caimp} in powers of the coupling $g_0^2$ and setting 
$S_{\rm h}^1=S_{\rm h}^{\rm EH}$, for which $c_{\rm A}^{\rm stat, (1)}$ is known 
from~\cite{castat}, we obtain
\begin{equation}
\castat = c_{\rm A}^{\rm stat, (1)} \,g_0^2 + \rmO(g_0^4)\,,\quad  
c_{\rm A}^{\rm stat, (1)}=0.0072(4) \,,\quad
 \;\,{\rm for} \; S_{\rm h}^{\rm s}\; {\rm and} \;S_{\rm h}^{\rm A}\;.
\label{caA}
\end{equation}
The ratios in \eq{caimp}
have then been evaluated,  with $S_{\rm h}^1=S_{\rm h}^{\rm A}$,
on the configurations generated in runs I to IV of table~\ref{param}
in order to obtain $\castat$ for 
$S_{\rm h}^2=S_{\rm h}^{\rm HYP1}, \; S_{\rm h}^{\rm HYP2}$.
There the size $L/a$ and the $\beta$ values have been chosen  such that $L$ is kept fixed to 
$1.436 r_0$ with $r_0$ from \cite{pot:r0,pot:r0_SU3}. 
Our results are shown in figure~\ref{ca_fig}.
%
%
The result for 
$S_{\rm h}^{\rm s}$ 
provides a test of the numerical procedure since it
is known analytically (\eq{caA}).
We see that by ascribing to the numerical result an error, 
which covers the spread of the 
points we find agreement with \eq{caA}. This suggests that here higher orders in 
$g_0^2$ contribute little to the cutoff effects of the ratios appearing in \eq{caimp}.
We estimate in the same way the errors for $\castat$ of $S_{\rm h}^{\rm HYP1}$
and $S_{\rm h}^{\rm HYP2}$, quoting the result
\begin{eqnarray}
&&\castat \approx 0.039(4)\,g_0^2  \,,\quad\;\;
 {\rm for} \; S_{\rm h}^{\rm HYP1}  \label{caHYP1}\\
&&\castat \approx 0.220(14)\,g_0^2  \,,\quad
 {\rm for} \; S_{\rm h}^{\rm HYP2} \;.\label{caHYP2}
\end{eqnarray}
Here the statistical errors are not the relevant ones. Rather our dominant
error is due to the assumption (tested to some extent
as just explained) that the 1-loop value for $\castat$
for action $S_{\rm h}^1=S_{\rm h}^{\rm A}$ is accurate for our values of $g_0^2$.
Apart from this, our determinations do provide a non-perturbative
computation of $\castat$ for the other actions. Thus
the errors quoted in eqs.~(\ref{caHYP1},\ref{caHYP2}) include a
reasonable (but not necessarily a safe) estimate of the perturbative
uncertainty.

Note that for $S_{\rm h}^{\rm HYP2}$ the improvement coefficient is
considerably larger than for the other actions. This means that before
improvement the linear $a$-effects are larger in this case and one might 
expect that this will be the case also for
higher order $a$-effects. We will see below that this is however not born 
out of our non-perturbative results.

Sensitivity to the improvement coefficient $\bastat$ is achieved by
exploiting the quark mass dependence of the correlation function
$\fastatimpr(x_0)$. Again, supposing $\bastat$ be known for the action
$S_{\rm h}^1$, we consider the improvement condition
\be
{\left.(1+a\bastat(S_{\rm h}^1)   m_{\rm q'}) \fastatimpr(T/2,S_{\rm
    h}^1)\right|_{m_{\rm q'}}
  \over \left.(1+a\bastat(S_{\rm h}^1)   m_{\rm q})
  \fastatimpr(T/2,S_{\rm h}^1)\right|_{m_{\rm q}}}  = 
{\left.(1+a\bastat(S_{\rm h}^2)   m_{\rm q'}) \fastatimpr(T/2,S_{\rm
    h}^2)\right|_{m_{\rm q'}}
  \over \left.(1+a\bastat(S_{\rm h}^2)   m_{\rm q})
  \fastatimpr(T/2,S_{\rm h}^2)\right|_{m_{\rm q}}} \;, 
\label{baimp}
\ee
with $\theta=1$, and solve for $\bastat$ of $S_{\rm h}^2$.

\begin{figure}[htb]
\hspace{0cm}
\centering
\epsfig{file=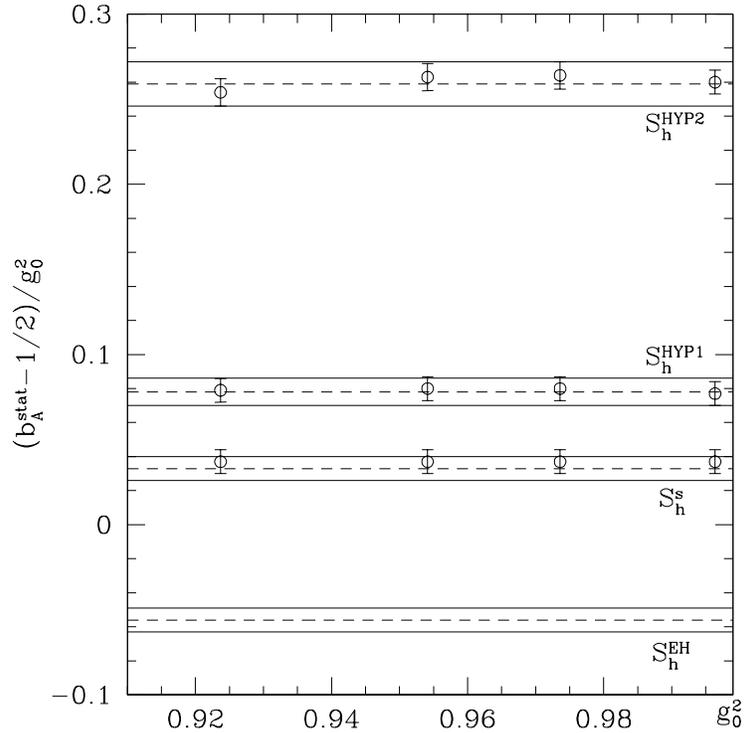,width=9.7cm,angle=0}
\caption{\footnotesize{\sl 
Numerical results for the improvement coefficient $\bastat$, 
presented as in \fig{ca_fig}.
}}
\label{ba_fig}
\end{figure}

In perturbation theory $\bastat$ has been computed to 1-loop order
in~\cite{zastat:pap1} for the Eichten--Hill regularization. Similarly to what
we did for $\castat$, we have then analytically computed $\bastat$ for
$S_{\rm h}^{\rm A}$, i.e. we have expanded~\eq{baimp} in perturbation
theory with $S_{\rm h}^{\rm 1}=
S_{\rm h}^{\rm EH}$ and $S_{\rm h}^{\rm 2}=S_{\rm h}^{\rm A}$.
In this computation we set $m_{\rm q}=0$ and, following~\cite{impr:babp},
$m_{\rm q'}$ such that
$Lm_{\rm R'}=0.24$, with $m_{\rm R'}$ the quark mass renormalized 
at scale $\mu=1/L$ in the
minimal subtraction scheme on the lattice. The result reads
\be
\bastat=1/2+  b_{\rm A}^{\rm stat, (1)} \,g_0^2 + \rmO(g_0^4)\,,\quad 
b_{\rm A}^{\rm stat, (1)}=0.033(7)  \,,\quad
 \;\,{\rm for} \; S_{\rm h}^{\rm s}\; {\rm and} \;S_{\rm h}^{\rm A}\;.
\label{baA}
\ee
Next we impose~\eq{baimp} on the data sets I to IV in table~\ref{param}.
Here, both $\kappa$ values and $\theta=1$ are used, where the second 
$\kappa$ value was determined such that $Lm_{\rm R'}=0.24$ 
{\em in the SF-scheme} at scale $\mu=1/(1.436 r_0)$.
With $S_{\rm h}^1=S_{\rm h}^{\rm A}$, we obtain
\begin{eqnarray}
&&\bastat \approx 1/2+0.078(8)\,g_0^2  \,,\quad\;\;
 {\rm for} \; S_{\rm h}^{\rm HYP1}  \label{baHYP1}\\
&&\bastat \approx 1/2+0.259(13)\,g_0^2  \,,\quad
 {\rm for} \; S_{\rm h}^{\rm HYP2} \;.\label{baHYP2}
\end{eqnarray}
The errors have been estimated as in the case of $\castat$. We show
the  numerical results in figure~\ref{ba_fig}.
Again the improvement coefficient turns out largest for
$S_{\rm h}^{\rm HYP2}$.
%
%

One comment is in order here. In the  condition in~\eq{baimp},
O$(am_{\rm q})$-terms  with coefficients
$b_{\rm g}$ and $b_{\zeta_{\rm h}}$~\cite{zastat:pap1}
are neglected, while
terms proportional to $b_{\zeta}$ (see~\cite{impr:pap5}) drop out. 
Of course,  
$b_{\zeta_{\rm h}}$ is O($g_0^4 \nf$) and therefore 
irrelevant in a 1-loop computation. While $b_g$ is O$(g_0^2 \nf)$, 
it enters expectation values only with an additional
factor $g_0^2$ (unless there is a non-zero background gauge field).
In other words, \eq{baimp} is correct up to $\rmO(g_0^4)$ in
full QCD, but in the quenched approximation it
is valid also non-perturbatively.
\subsection{Scaling tests}
On the same set of configurations used for the determination of 
$\castat$ and $\bastat$ (runs I--IV), and for $\kappa=\kappa_{\rm c}$,  we 
have computed (for each regularization) the ratios
\begin{equation}
\xi_{\rm A}(\theta,\theta')={{\left.\fastatimpr(T/2)\right|_{\theta}}\over
{\left.\fastatimpr(T/2)\right|_{\theta'}}} \; , \quad
\xi_1(\theta,\theta') = {{\left. f_1^{\rm stat} \right|_{\theta}}\over
{\left. f_1^{\rm stat} \right|_{\theta'}}} \; , 
\quad h(d/L)={{f_1^{\rm hh}(d)}\over{f_1^{\rm hh}(L/2)}}\; .
\label{xi_del_h}
\end{equation}
Here and in the following $\fastatimpr(x_0)$ refers to the 
1-loop improved axial current in \eq{e_aimpr}. The correlation function 
$f_1^{\rm stat}$  is schematically represented in the right part of 
\fig{fig_fA}, it precisely reads
\begin{equation}
f_1^{\rm stat}=- {{a^{12}}\over{2L^6}} \sum_{\bf u,v,y,z}  \;  \langle 
\zetalbprime ({\bf u}) 
\gamma_5 
\zetahprime ({\bf v}) \; \zetahb ({\bf y}) \gamma_5 
\zeta_{\rm l} ({\bf z}) \rangle
\; ,
\end{equation} 
with the primed fields living on the boundary at $x_0=T$.
The quantity $h(d/L)$ was introduced in~\cite{zastat:pap3}. It is 
defined through the boundary to boundary SF correlator
\begin{equation}
f_1^{\rm hh}(x_3)= -{{a^8}\over{2L^2}} \sum_{x_1,x_2,{\bf y},{\bf z}}
 \; \langle \zetahbprime ({\bf x}) 
\gamma_5 
\zetahprime ({\bf 0}) \; \zetahb ({\bf y}) \gamma_5 
\zetah ({\bf z}) \rangle \; ,
\label{f1hh}
\end{equation}
note that the sum runs on $x_1$ and $x_2$ and therefore yields an 
$x_3$-dependent correlation function. 

\begin{figure}[htb]
\hspace{0cm}
\centering
\epsfig{file=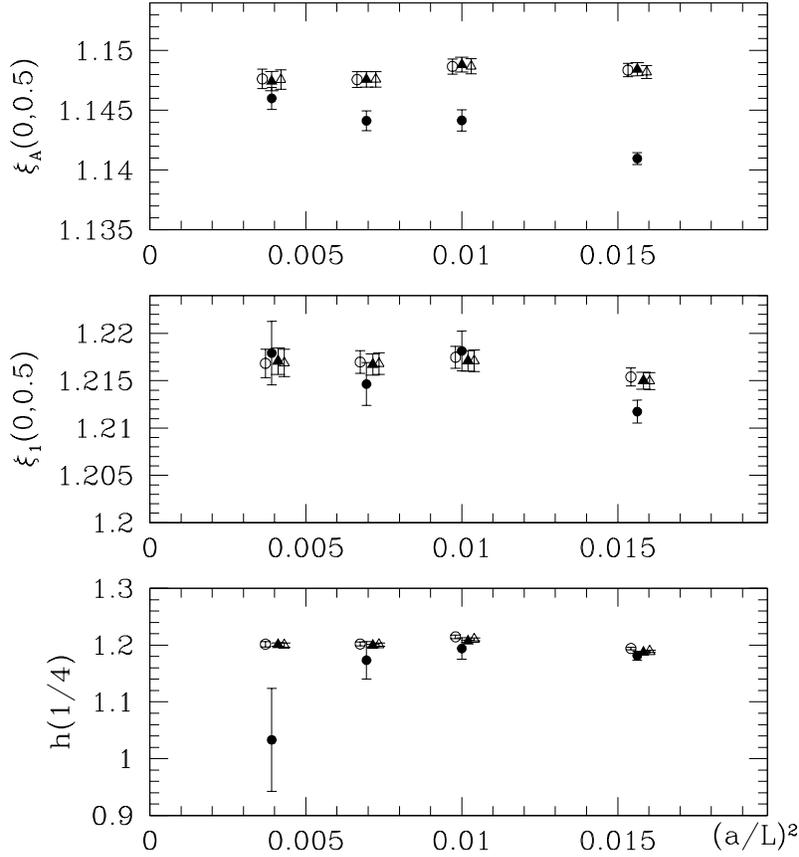,width=10.5cm,angle=0}
\caption{\footnotesize{\sl 
Scaling plots for $\xi_{\rm A}(0,0.5)$, $\xi_1(0,0.5)$ and $h(1/4)$.
Symbols are as in figure~\ref{f_rns}, some of them have been shifted on the
 horizontal axis for clarity. Again the results for $ S_{\rm h}^{\rm s}$
are similar to the ones from $S_{\rm h}^{\rm A}$ and haven't been plotted.
}}
\label{fig_scal}
\end{figure}

All the quantities in~\eq{xi_del_h} have a
finite continuum limit in a fixed, finite, volume. 
The continuum result is 
universal, but the way it is approached depends on the details of the 
regularization. The scaling of these ratios can then provide an 
idea about the size of discretization effects for the static actions in 
eqs.~(\ref{SAPE}~-~\ref{SHYP}). Our results are shown in figure~\ref{fig_scal}.
%
%
The cutoff effects in $\xi_{\rm A}$ are somewhat larger with the 
Eichten--Hill action than with any of our alternatives. Still,
it is more relevant to note that
all cutoff effects visible in the figure are
very small and linear in $a^2$. This suggests 
that the O($a$) 
improvement programme has been implemented in a satisfactory way for the lattice
spacings considered here -- also for $S_{\rm h}^{\rm HYP2}$, where the improvement
coefficients turn out to be not that small.

The gain in statistical precision brought by the new discretizations of the static
action can also be seen in figure~\ref{fig_scal}, especially for the quantity
$h(1/4)$, which involves two static quarks propagating over the whole temporal 
extent.

\begin{figure}[htb]
\hspace{0cm}
\centering
\epsfig{file=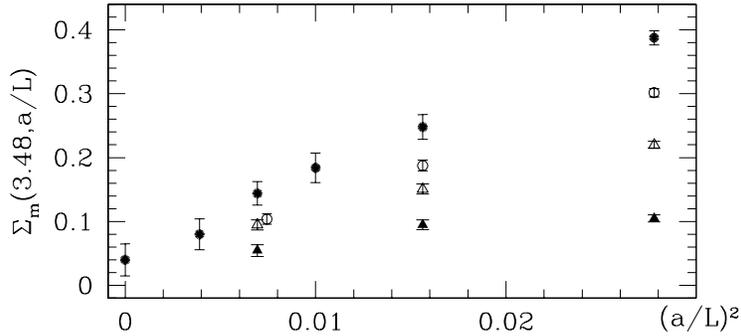,width=9.7cm,angle=0}
\caption{\footnotesize{\sl 
The step scaling function $\Sigmam$ for the different static actions.
Symbols are chosen as in \protect\fig{f_rns}. Results for the 
Eichten--Hill action are taken from \protect\cite{hqet:pap1}.
}}
\label{f:sigmam}
\end{figure}

Another interesting observable is the step scaling function
introduced in \cite{lat01:rainer} for the renormalization
of the b-quark mass and further discussed in the following section. 
Here we simply define it as
\bes
\Sigma_{\rm m}(u,a/L)&=&2L\left[\Gamma_{\rm stat}(2L)-\Gamma_{\rm stat}(L)
        \right]_{u=\gbar^2(L)}\; \\
 \meffstat(L) &=& {{1}\over{2a}} \ln \left[\fastatimpr(x_0-a)/\fastatimpr(x_0+a) 
   \right]\;,
 \quad \theta=1/2\,,\;T=L\,,
 \label{e:meffstat}
\ees 
in terms of the correlation function $\fastatimpr$ and 
the Schr\"odinger functional 
coupling  $\gbar(L)$ \cite{SF:LNWW,alpha:SU3,mbar:pap1},
which fixes the length scale $L$.
The continuum limit of $\Sigmam$ is, for each value of the coupling $u$, 
a universal quantity,
independent of the regularization used. The step scaling function
can hence be used for a further scaling study, in particular
since discretization errors with the EH action turn out 
to be clearly visible \cite{hqet:pap1}. 
\Fig{f:sigmam} shows that they are rather linear in $(a/L)^2$ for
all actions considered. Compared to the EH action, the alternative
ones show smaller lattice artefacts. In fact they are the smallest
for $S_{\rm h}^{\rm HYP2}$ which also showed the best noise to signal ratio.
We will return to another scaling test 
in \sect{s:ssf}.

\section{Renormalization of the axial current and quark mass \label{s:renorm}}
In the effective theory the static--light axial current is not derived
from a symmetry transformation of the action. The renormalized current
is therefore scale dependent. This dependence has been studied 
non-perturbatively, over a wide energy range, in the SF scheme
(see~\cite{zastat:pap3}). The main quantity considered in that study is 
the step scaling function $\Sigma_{\rm A}^{\rm stat}(u,a/L)$  defined as
\be
\Sigma_{\rm A}^{\rm stat}(u,a/L)=
\left.{{\zastat(g_0,2L/a)}\over{\zastat(g_0,L/a)}}\right|_{\gbar^2(L)=
u,\;m_{\rm q}=0} \;,
\label{ssfA}
\ee
with 
\be
\zastat={{\Xi^{(0)}}\over{\Xi}}\;, \quad {\rm and} \quad
\Xi=\left.{{\fastatimpr(L/2)}\over{\left[f_1\; f_1^{\rm hh}(L/2)
      \right]^{1/4}}}\right|_{\theta=0.5} \;, 
\label{newZ}
\ee
here $\Xi^{(0)}$ is the tree-level value of $\Xi$, as in~\cite{zastat:pap3}. In~\eq{newZ} $f_1$
is the correlator between two light-quark pseudoscalar boundary
sources
\be
f_1=- {{a^{12}}\over{2L^6}} \sum_{\bf u,v,y,z}  \;  \langle
\zetabarprime_1 ({\bf u})
\gamma_5
\zeta\kern1pt'_2 ({\bf v}) \; \zetabar_2 ({\bf y}) \gamma_5
\zeta_1 ({\bf z}) \rangle
\; ,
\ee
and we remind the reader that $\gbar(L)$ is the Schr\"odinger functional 
coupling. In the first part of this section we report on
a scaling study of $\Sigma_{\rm A}^{\rm stat}$.

In addition, for a chosen reference scale $L_{\rm ref}=1.436 r_0$, we
are interested in the dependence of the renormalization factor
$\zastat(g_0,L_{\rm ref}/a)$ on the bare coupling $g_0$. This allows to
match bare matrix elements of the axial current to continuum ones (up
to cutoff effects). The renormalization factor itself clearly depends
on the regularization, and needs to be recomputed for the actions 
in eqs.~(\ref{SAPE}~-~\ref{SHYP}). 

In the quenched approximation the b-quark mass has been obtained at
leading order in HQET through the matching of the effective theory
to the full one~\cite{lat01:rainer,hqet:pap1}. The mass of the B-meson is used
in the procedure as phenomenological input. Therefore, although the
matching is performed in a small volume, the effective theory has then
to be connected to large volumes. Improving on this part of the
computation would reduce the error on the result for the b-quark mass.
The use of the static actions introduced here (in place of the
Eichten-Hill action used in~\cite{lat01:rainer,hqet:pap1}) has been proven to
be very efficient in this respect (see~\cite{reviews:physcoll}, where $S_{\rm
  h}^{\rm HYP1}$ has been considered). In the last part of this
section we briefly discuss the strategy which has been used to compute
the b-quark mass, with
particular emphasis on the quantities we recompute in the new
discretizations.
\subsection{Results for the step scaling function and $\zastat$ \label{s:ssf}}
\begin{figure}[htb]
\hspace{0cm}
\centering
\epsfig{file=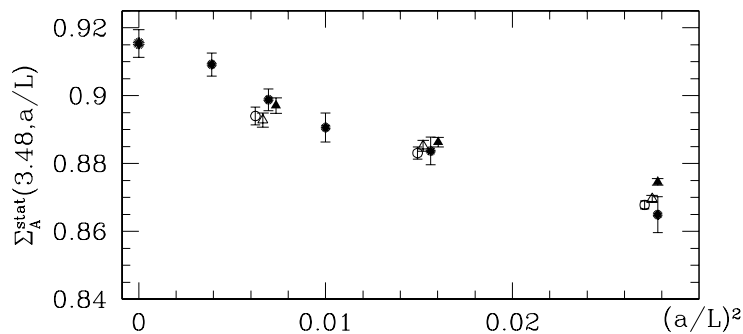,width=9.7cm,angle=0}
\caption{\footnotesize{\sl
The step scaling function $\Sigma(3.48,a/L)$. 
Symbols are as in \protect\fig{f_rns}, 
some of them have been shifted on the
 horizontal axis for clarity. The results for $ S_{\rm h}^{\rm s}$
are similar to those from $S_{\rm h}^{\rm A}$ and haven't been
included in the plot. Data for the Eichten-Hill action and the
continuum limit value have been taken from~\cite{zastat:pap3}. 
}}
\label{ssf3}
\end{figure}

The step scaling function in~\eq{ssfA} has been evaluated for $u=3.48$ and
with lattice resolution $L/a=6,8,12$  
on the configurations produced in the runs
ZI--sZIII of table~\ref{param} (the same sets were already used
in \fig{f:sigmam}). In figure~\ref{ssf3}
the results are compared to those for the Eichten-Hill action,
obtained in~\cite{zastat:pap3}. In this case a larger set of lattice spacings
was used  and the data have been extrapolated to the continuum limit.
Here again we see that the different regularizations yield very similar results
already for finite lattice spacing, although in principle agreement needs to be
found only after having taken the continuum limit.
\begin{table}[htb]
\centering
\begin{tabular}{c|cccc}
\hline\\[-2.5ex]
  & \multicolumn{4}{c}{$\castat=0$} \\[1ex]  
$\beta$  &$S_{\rm h}^{\rm A}$ &$S_{\rm h}^{\rm s}$   &$S_{\rm h}^{\rm HYP1}$ &$S_{\rm h}^{\rm HYP2}$ \\[1ex]
\hline 
    6.0219 & 0.7873(9)  & 0.7880(8)   &  0.8015(7)  &   0.8823(7)   \\
    6.1628 & 0.7722(10) & 0.7727(10)  &  0.7845(9)  &   0.8579(9)   \\
    6.2885 & 0.7633(10) & 0.7640(10)  &  0.7746(9)  &   0.8409(10)  \\ 
    6.4956 & 0.7565(12) & 0.7569(12)  &  0.7653(11) &   0.8227(11)   
\\[1ex]
\hline\\[-2.5ex] 
&\multicolumn{4}{c}{$\castat$=1-loop} \\[1ex]
$\beta$ &$S_{\rm h}^{\rm A}$ &$S_{\rm h}^{\rm s}$   &$S_{\rm h}^{\rm HYP1}$ &$S_{\rm h}^{\rm HYP2}$ \\[1ex]
\hline
    6.0219 & 0.7862(8)  & 0.7868(8)   &  0.7933(7)  &   0.8406(7)  \\ 
    6.1628 & 0.7708(10) & 0.7714(10)  &  0.7770(9)  &   0.8196(8)  \\
    6.2885 & 0.7619(10) & 0.7624(10)  &  0.7669(9)  &   0.8054(8)  \\
    6.4956 & 0.7543(13) & 0.7547(13)  &  0.7580(12) &   0.7913(11) \\[1ex]
\hline
\end{tabular}
\caption{\footnotesize{\sl Results for $\zastat$ at the scale $L_{\rm
      ref} =1.436 r_0$.
}}\label{Zatable}
\end{table}

At the scale $L_{\rm ref}=1.436 r_0$ we have also computed the
renormalization constant $\zastat$ in the $\beta$-range relevant for large volume
simulations (again data from runs I-IV have been used here). We summarize the 
results in table~\ref{Zatable}.
For $6 \leq \beta \leq 6.5$ we parameterize them in the form
$(x=\beta-6)$
\begin{eqnarray}
\zastat(g_0,L_{\rm ref}/a)&=&0.7889-0.1290\;x+0.1197\;x^2 
\;,\quad {\rm for} \; S_{\rm h}^{\rm A} 
\label{zaAI}\\
\zastat(g_0,L_{\rm ref}/a)&=&0.7895-0.1291\;x+0.1189\;x^2 
\;,\quad {\rm for} \; S_{\rm h}^{\rm s} \\
\zastat(g_0,L_{\rm  ref}/a)&=&0.7962-0.1368\;x+0.1207\;x^2 
\;,\quad {\rm for} \; S_{\rm h}^{\rm HYP1} \\
\zastat(g_0,L_{\rm ref}/a)&=&0.8443-0.1740\;x+0.1351\;x^2 
\;,\quad {\rm for} \; S_{\rm h}^{\rm HYP2} 
\label{zaHYP2I}
\end{eqnarray}
when $\castat$ is set to its 1-loop value, and
\begin{eqnarray}
\zastat(g_0,L_{\rm ref}/a)&=&0.7900-0.1286\;x+0.1231\;x^2 
\;,\quad {\rm for} \; S_{\rm h}^{\rm A} \\
\zastat(g_0,L_{\rm ref}/a)&=&0.7907-0.1287\;x+0.1224\;x^2 
\;,\quad {\rm for} \; S_{\rm h}^{\rm s} \\
\zastat(g_0,L_{\rm  ref}/a)&=&0.8044-0.1404\;x+0.1244\;x^2 
\;,\quad {\rm for} \; S_{\rm h}^{\rm HYP1} \\
\zastat(g_0,L_{\rm ref}/a)&=&0.8866-0.1995\;x+0.1423\;x^2 
\;,\quad {\rm for} \; S_{\rm h}^{\rm HYP2} 
\label{zaHYP2}
\end{eqnarray}
for $\castat=0$. The formulae reproduce the numbers in
table~\ref{Zatable} within their errors. They may be assigned an uncertainty
of about 2\textperthousand.

Given now a bare matrix
element $\Phi_{\rm bare}$ of the static--light axial current, the corresponding
matrix element $\Phi(\mu)$ renormalized in the SF scheme at the scale 
$\mu^{-1}=1.436r_0$ can be written
\be
\Phi(\mu)=\zastat(g_0,L_{\rm ref}/a)\;\Phi_{\rm bare}(g_0)\;, \quad
\mu^{-1}=L_{\rm ref} \;,
\ee
with $\zastat(g_0,L_{\rm ref}/a)$ from eqs.(\ref{zaAI}-\ref{zaHYP2}) 
depending on the regularization used to compute the bare matrix element.
Finally the Renormalization Group Invariant (RGI) matrix element
$\Phi_{\rm RGI}$ reads
\be
\Phi_{\rm RGI} =  {{\Phi_{\rm RGI}}\over{\Phi(\mu)}} \; \Phi(\mu) \;,
\ee
where we have factorized out the universal ratio $\Phi_{\rm RGI}/
\Phi(\mu)$, which has been computed in the continuum limit
in~\cite{zastat:pap3}. There one finds the result
\be
{{\Phi(\mu)}\over{\Phi_{\rm RGI}}}=1.088(10)\;, \quad {\rm at}
\quad \!\!  \mu=(1.436 r_0)^{-1}\;, 
\label{unifact}
\ee
which, as discussed, applies also to the discretizations introduced here.
Moreover, as the error on the ratio in~\eq{unifact} refers to a continuum result, 
it propagates into $\Phi_{\rm RGI}$ only once $\Phi(\mu)$ has been extrapolated to 
the continuum limit.
\subsection{Renormalization of the quark mass}
In~\cite{hqet:pap1} a strategy for the non-perturbative 
computation of the b-quark mass 
in the static approximation has been introduced and discussed. We refer to that publication
for all the details of the method and we only remind the reader of the basic formula, from
which the RGI b-quark mass $M_{\rm b}$ can be implicitly derived:
\be
L_0 m_{\rm B}=L_0 \Gamma(L_0, M_{\rm b}) + \sum_{k=0}^{K-1}2^{-(k+1)} \sigma_{\rm m}(u_k)
+L_0 \Delta E \;,
\label{basic_b}
\ee
where the use of a finite volume scheme like the Schr\"odinger functional is assumed.
In~\eq{basic_b}
\begin{itemize}
\item $L_0$ is the linear extent of the small volume where the matching between HQET and QCD is performed.
\item $m_{\rm B}$ is the physical (spin-averaged) ${\rm B}_{({\rm s})}$-meson mass.
\item $\Gamma$ is an energy defined in terms  of heavy-light correlators
(with a heavy quark of mass $M$) in a finite volume $L^3$. A relevant example is
\bes
\Gamma(L,M)&=& \frac14(\meffp+3\meffv)\,,\quad
 \meffp = {{1}\over{2a}} \ln \left[ f_{\rm A}(x_0-a)/f_{\rm A}(x_0+a) \right]\;,
\ees
($
{{x_0}/{T}}=1/2 \; {\rm with}\; {{T}/{L}}\; {\rm fixed}
$)
where $\meffv$ is defined analogously in the vector channel.
Their static version $\Gamma_{\rm stat}(L)$ was defined in
\eq{e:meffstat}.
Its infinite volume limit gives the 
static binding energy $E_{\rm stat}$ 
in~\eq{QMrep}.
\item $\sigma_{\rm m}$ is a step scaling function
\be
\sigma_{\rm m}(\bar{g}^2(L))=2L\left[\Gamma_{\rm stat}(2L)-\Gamma_{\rm stat}(L)\right]\; .
\ee 
Notice that in the difference the dependence of 
$\Gamma_{\rm stat}$ on $\delta m_{\rm W}$ cancels.
Once the effective theory and the full one are matched in small volume
the step scaling function provides, within the effective theory, the connection to large 
volumes, where contact with phenomenology can be made. This large scale difference is 
covered in several steps, such that at each stage scales differing only by a factor two
appear. In this sense the sum in~\eq{basic_b} connects the size $L_0$ to $L_K=2^KL_0$,
therefore $u_k=\bar{g}^2(2^kL_0)$ in the equation.
\item $\Delta E$ is an energy difference 
\be
\Delta E =E_{\rm stat}-\Gamma_{\rm stat}(L_K)\;.
\ee
\end{itemize}
It is important to remark that in the way~\eq{basic_b} has been written all the quantities
on the r.h.s. are independent of $\delta m_{\rm W}$ and can be computed in the continuum limit. 

In real applications of the method (see~\cite{lat01:rainer,hqet:pap1}) 
the binding energy $E_{\rm stat}$ has been computed on a lattice
of size 1.6~fm while for $L_K$ it sufficed to have almost half of it, more precisely
$L_K=L_{\rm ref}=1.436r_0$. In view of combining this strategy with the use of  
the new static actions
 we have calculated the quantity $a\Gamma_{\rm stat}(g_0,L_{\rm ref}/a)$, 
on the configurations generated in runs I-IV of table~\ref{param}.
The results are collected in table~\ref{Gammatable}.
\begin{table}[htb]
\centering
\begin{tabular}{c|cccc}
\hline\\[-2.5ex]
  & \multicolumn{4}{c}{$\castat=0$} \\[1ex]  
$\beta$  &$S_{\rm h}^{\rm A}$ &$S_{\rm h}^{\rm s}$   &$S_{\rm h}^{\rm HYP1}$ &$S_{\rm h}^{\rm HYP2}$ \\[1ex]
\hline \\[-2.5ex]
    6.0219 &  0.5868(5)  & 0.5000(4)   &  0.2044(4)  &   0.1744(4)  \\ 
    6.1628 &  0.5575(4)  & 0.4763(4)   &  0.1921(4)  &   0.1625(4)  \\
    6.2885 &  0.5326(3)  & 0.4558(3)   &  0.1812(3)  &   0.1537(3)  \\
    6.4956 &  0.4948(3)  & 0.4243(3)   &  0.1643(3)  &   0.1390(3)
\\[1ex]
\hline\\[-2.5ex] 
&\multicolumn{4}{c}{$\castat$=1-loop} \\[1ex]
$\beta$ &$S_{\rm h}^{\rm A}$ &$S_{\rm h}^{\rm s}$   &$S_{\rm h}^{\rm HYP1}$ &$S_{\rm h}^{\rm HYP2}$ \\[1ex]
\hline\\[-2.5ex]
    6.0219 &  0.5871(5)  & 0.5003(4)   &  0.2050(4)  &   0.1769(4)  \\ 
    6.1628 &  0.5576(4)  & 0.4763(4)   &  0.1923(4)  &   0.1641(4)  \\
    6.2885 &  0.5335(3)  & 0.4567(3)   &  0.1822(3)  &   0.1547(3)  \\
    6.4956 &  0.4958(3)  & 0.4254(3)   &  0.1653(3)  &   0.1396(3) \\[1ex]
\hline
\end{tabular}
\caption{\footnotesize{\sl Results for $a\Gamma_{\rm stat}(g_0,L_{\rm ref}/a)$
 with $L_{\rm ref} =1.436 r_0$, $\beta=6/g_0^2$.
}}\label{Gammatable}
\end{table} 

The data can be described within one standard deviation by polynomial expressions.  Explicitly, 
for the 1-loop value of $\castat$ and for $ 6 \leq \beta=6/g_0^2 \leq 6.5 $ we provide the parameterizations
$(x=\beta-6)$
\begin{eqnarray}
a\Gamma_{\rm stat}(g_0,L_{\rm ref}/a)&=&
0.5916-0.2140\;x +0.0418\;x^2
\,,\quad {\rm for} \; S_{\rm h}^{\rm A} \\
a\Gamma_{\rm stat}(g_0,L_{\rm ref}/a)&=&
0.5040-0.1726\;x +0.0285\;x^2
\,,\quad {\rm for} \; S_{\rm h}^{\rm s} \\
a\Gamma_{\rm stat}(g_0,L_{\rm ref}/a)&=&
0.2068-0.0890\;x +0.0106\;x^2
\,,\quad{\rm for} \; S_{\rm h}^{\rm HYP1} \\
a\Gamma_{\rm stat}(g_0,L_{\rm ref}/a)&=&
0.1787-0.0915\;x +0.0255\;x^2
\,,\quad {\rm for} \; S_{\rm h}^{\rm HYP2}
\end{eqnarray}
with an uncertainty of about 2.5\textperthousand.
In addition we see from the table that $\castat$ has very little impact on this 
quantity. 


\section{Summary and conclusions}
We have proposed and investigated alternative discretizations
for static quarks on the lattice. All of them have automatic
$\rmO(a)$ improvement, i.e. energies approach the continuum
limit with correction $\rmO(a^2)$ if the light-quark sector 
is improved. The purpose of introducing these actions was
to achieve a better noise to signal ratio at large Euclidean times.
This goal could be reached, since the exponential growth
of the noise to signal ratio is non-universal and can be reduced 
by choosing a discretization with a small self energy for the 
static quark (relative to the lattice potential at the origin).
The two actions with HYP links are best in this respect. 

Of course a prime criterion for choosing an action is to have 
small lattice artifacts. We have therefore investigated several
quantities
(figures \ref{fig_scal},\ref{f:sigmam},\ref{ssf3}). 
It turns out that lattice artifacts
are mostly indistinguishable for the different actions 
except for the case of $\Sigmam$, \fig{f:sigmam} (and to 
a smaller extent $\xi_{\rm A}$), where the
lattice artifacts with the two actions with HYP links are comfortably
small compared to the ones with the standard action. 
Since an implementation of two or three static actions 
is no problem in any practical computation, 
it is advisable to compute with both $S_{\rm h}^{\rm HYP2}$ and
$S_{\rm h}^{\rm HYP1}$ and to perform continuum extrapolations 
constrained by universality. This promises to stabilize
continuum extrapolations in a non-trivial manner, since 
e.g. \fig{f:sigmam} provides an example where the $a$-effects
are different for the two actions. 

In view of future applications we have computed all coefficients
needed to improve and renormalize the static-light axial current
for all actions considered. Here, the renormalization factors
are now known non-perturbatively (in the quenched approximation)
and the improvement coefficients are given up to 
uncertainties $\rmO(g_0^4)$. In the course of these computations we
again observed that the $a$-effects (now the ones before $\rmO(a)$
improvement) are different for the different actions.

We finally emphasize that in \cite{pot:intermed}
it was shown that the force between static quarks, computed e.g.
through Polyakov loop correlation functions, is automatically
$\rmO(a)$ improved. This is derived from the corresponding
property of the Eichten Hill action for static quarks. 
This proof is also valid when the alternative actions are employed.
Of course a significant gain in the noise to signal ratio can be achieved
in this way as has first been seen in \cite{HYP:pot}. Such a gain
in precision is especially
relevant in the theory with dynamical quarks, where the noise-reduction methods
of \cite{algo:multilevel} are not applicable. Indeed, 
with the action $S_{\rm h}^{\rm A}$, 
it has been possible very recently to observe string breaking
in $\nf=2$ QCD \cite{pot:sesam}.

In addition, these discretizations can be used to efficiently calculate the $1/\mbeauty$ 
corrections to the static limit. In this context a preliminary study has been
presented in~\cite{lat04:stephan}.

\vspace{0.4cm}
{\bf Acknowledgements.} We are grateful to H.~Simma and F.~Knechtli for useful 
discussions. 
Many thanks go to R. Hoffmann and F. Knechtli for performing the computation 
of $\eselfone$ for the $S_{\rm h}^{\rm HYP}$ actions. 
We thank DESY for allocating computer time on the APEmille
computers at DESY Zeuthen to this project and the APE group for its help. This
work is also supported by the EU IHP Network on Hadron Phenomenology from 
Lattice 
QCD under grant HPRN-CT-2000-00145 and by the Deutsche Forschungsgemeinschaft 
in the SFB/TR 09.

\begin{appendix}
\section{Perturbation theory for the APE action \label{s:PT}}

\subsection{Feynman rules}

\begin{table}[tb]
\begin{center}
\begin{tabular}{c @{\qquad\qquad} c @{\qquad\qquad} c 
@{\qquad\qquad} c}
\hline\\[-1ex]
$i$ & $\mu(i)$ & $s(i)$ & $h_i^{(1)}(\bbp)$ \\[1ex]
\hline\\[-1ex]
$0$ & $0$ & $0$  & $1-\frac{\alpha}{6} \sum_{k=1}^3 a^2 \phat_k^2$ \\[1ex]
$1,2,3$ & $i$ & $0$  & ${\rm i}\frac{\alpha}{6} a \phat_{\mu(i)} $ \\[1ex]
$4,5,6$ & $i-3$ & $1$  & $-{\rm i}\frac{\alpha}{6} a \phat_{\mu(i)}$ \\[1ex]
\hline
\end{tabular}
\caption{\footnotesize{\sl
The heavy-quark gluon vertex.}}
\label{table:vertex1}
\end{center}
\end{table}
\begin{table}[tb]
\begin{center}
\begin{tabular}{c @{\qquad} c @{\qquad} c @{\qquad} c 
@{\qquad} c @{\qquad} c}
\hline\\[-1ex]
$i$ & $\mu(i)$ & $\nu(i)$ & $s(i)$ & $t(i)$ & $h_i^{(2)}(\bbp,\bq)$ \\[1ex]
\hline\\[-1ex]
$0$ & $0$ & $0$ & $0$ & $0$ & $1-\frac{\alpha}{6} \sum_{k=1}^3 \sin^2[\frac{a(p_k+q_k)}{2}]$ \\[1ex]
$1,2,3$ & $i$ &  $i$ &$0$ & $0$ & $\frac{\alpha}{3} \cos[\frac{a(p_{\mu(i)}+q_{\mu(i)})}{2}]$ \\[1ex]
$4,5,6$ & $i-3$ & $i-3$ & $1$ & $1$ & $\frac{\alpha}{3} \cos[\frac{a(p_{\mu(i)}+q_{\mu(i)})}{2}]$ \\[1ex]
$7,8,9$ & $i-6$ & $i-6$ & $0$ & $1$ & $-2 \frac{\alpha}{3} \cos[\frac{a(p_{\mu(i)}+q_{\mu(i)})}{2}]$ \\[1ex]
$10,11,12$ & $i-9$ & $0$ & $0$ & $0$ & 
 $2{\rm i} \frac{\alpha}{3} \sin[a(\frac{p_{\mu(i)}}{2}+q_{\mu(i)})]$ \\[1ex]
$13,14,15$ & $0$ & $i-12$ & $0$ & $1$ & 
 $-2{\rm i} \frac{\alpha}{3} \sin[a(p_{\nu(i)}+\frac{q_{\nu(i)}}{2})]$ \\[1ex]
\hline
\end{tabular}
\caption{\footnotesize{\sl
The 4 point heavy-quark gluon vertex.}
\label{table:vertex2}}
\end{center}
\end{table}
We use the conventions of \cite{zastat:pap1,impr:pap1}. Here we generalize the
Feynman rules given for the EH action in appendix~B.1 of \cite{zastat:pap1}
(again we use the notation of that reference), for the
action 
\be
\label{SAPEalpha}
S_{\rm h}^{\rm A,\alpha} : \quad W^{\rm A,\alpha}(x,0) = (1-\alpha)U(x,0) +
	\alpha V(x,0) \; ,
\ee
which reduces to \eq{SAPE}, when $\alpha=1$. We set $\delta m_{W}=0$.
The propagation of a static quark from the boundary of the SF to the bulk is
given by the matrix $\Hheavy$ with a
perturbative expansion 
\be
H_{\rm h}(x) = \sum_{k=0}^{\infty} g_0^{k}H^{(k)}_{\rm h}(x) \,.
\label{eq:Hh}
\ee
The terms proportional to $g_0^k$ with $k=0,1,2$ are given by
\bes
H^{(0)}_{\rm h}(x) &=& P_+\; ,
\\
H^{(1)}_{\rm h}(x) &=& - \frac{a}{L^3} \sum_{\bbp} \sum_{u_0 = a}^{x_0}
{\rm e}^{{\rm i} \bbp \bx} 
\sum_{i=1}^{6} h^{(1)}_i(\bbp) \tilde q^a_{\mu(i)}(u_0 - a + a~s(i);\bbp) T^a
P_+\; ,
\\
H^{(2)}_{\rm h}(x) &=& H^{(2)}_{\rm h}(x)_1 + H^{(2)}_{\rm h}(x)_2\; ,
\\
H^{(2)}_{\rm h}(x)_1 &=& \frac{a^2}{L^6} \sum_{\bbp,\bq} \sum_{u_0 = a}^{x_0} \sum_{v_0=a}^{u_0-a}
{\rm e}^{{\rm i} (\bbp + \bq)\bx} \sum_{i=0}^{6} h^{(1)}_i(\bbp) \tilde q^a_{\mu(i)}(u_0 - a + a~s(i);\bbp) \\
&& \sum_{j=0}^{6} h^{(1)}_j(\bq) \tilde q^b_{\mu(j)}(v_0 - a + a~s(j);\bq) T^a
T^b P_+\; ,
\\
H^{(2)}_{\rm h}(x)_2 &=& \frac{a^2}{2L^6} \sum_{\bbp,\bq} \sum_{u_0 = a}^{x_0} 
{\rm e}^{{\rm i} (\bbp + \bq)\bx} \sum_{i=0}^{15} h^{(2)}_i(\bbp,\bq) \tilde q^a_{\mu(i)}(u_0 - a + a~s(i);\bbp) \\
&& \tilde q^b_{\nu(i)}(u_0 - a + a~t(i);\bq) T^a T^b P_+\; ,
\eea
where the gluon field in the time momentum representation is defined by
\bea
q_0(x) &=& \frac{1}{L^3}\sum_{\bbp} {\rm e}^{i \bbp \bx} \tilde{q}_0(x_0,\bbp), \\
q_k(x) &=& \frac{1}{L^3}\sum_{\bbp} {\rm e}^{i( \bbp \bx + ap_k/2)} \tilde{q}_k(x_0,\bbp)
\quad {\rm with~~} k = 1,2,3\; .
\eea 
The vertex functions
$h_{i}^{(1)}$ and $h_{i}^{(2)}$ 
are listed in tables~\ref{table:vertex1} and \ref{table:vertex2}. 
They are linear in $\alpha$ and reduce to the ones for the EH action
for $\alpha=0$. It is thus sufficient to give results for 
$\alpha=1$, to which we specialize from now on.

\subsection{Self energy} 

Here we compute, at one loop order, the linearly divergent contributions to the
static propagator and the potential at zero distance for the action 
$S_{\rm h}^{\rm A}$. 
We expand the correlation function
$f_1^{\rm hh}(x_3)$ defined in eq. (\ref{f1hh}) 
\be
{1 \over N_c}\fonehh(x_3) = 1 +
g_0^2\fonehhoneloop\Big(x_3,{a \over L}\Big)\; +\ldots.
\label{f1hh1l}
\ee
The Feynman diagrams that contribute to $\fonehhoneloop$ are given in fig. B.1
of \cite{zastat:pap3}. We then define
\be \label{e:eselfsf}
\eselfone = - \lim_{{a \over L} \rightarrow 0}
\fonehhoneloop\Big(x_3,{a \over L}\Big){a \over 2L}\; .
\ee

As explained in sec. 2.1 the relevant quantity for the signal to noise ratio
is $\Eself - V(0)/2$. 
In order to compute the divergent contribution of $V(0)$ we introduce
\be
\hat{f}_1^{\rm hh}(\bx)= -{{a^6}\over{2}} \sum_{{\bf y},{\bf z}}
 \; \langle \zetahbprime ({\bf x}) 
\gamma_5 
\zetahprime ({\bf 0}) \; \zetahb ({\bf y}) \gamma_5 
\zetah ({\bf z}) \rangle \; ,
\label{f1hhV0}
\ee
which is related to ${f}_1^{\rm hh}$ via
${f}_1^{\rm hh}(x_3)=(a/L)^2\sum_{x_1,x_2}\hat{f}_1^{\rm hh}(\bx)
$.
With the by now familiar notation for the one-loop coefficients,
we then define 
\be
\delta V^{(1)}(0) = -
\lim_{{a \over L} \rightarrow 0} \fonehhhoneloop\Big({\bf x}={\bf 0}
,{a \over L}\Big){a \over L}\; .
\ee
The combination
\be
r^{(1)} = \eselfone - {\delta V^{(1)}(0) \over 2}\; 
\ee
is listed in table~\ref{deltam}.

While the Feynman rules given above are valid for 
$S_{\rm h}^{\rm A}$, the result for $r^{(1)}$ is the same 
for the action $S_{\rm h}^{\rm s}$.
These two actions differ only by the normalization term
\be
t =  \left[ {{g_0^2}\over{5}}+\left(
{{1}\over{3}} \tr V^{\dagger}(x,0)V(x,0) \right)^{1/2} \right]^{-1}\; ,
\label{norm}
\ee
multiplying the links $W$. 
This normalization factor is gauge invariant. Consequently, at one-loop 
order, the gluon fields which are obtained from the expansion of 
$t$, are connected only to themselves. There is no connected graph 
from $t$ to the rest of the variables. In other words, these are
tadpole graphs, which factorize (at one-loop!) in the sense that 
\bes
  \label{e:Ot}
  \langle {\cal O}_{\rm h} \rangle_t = \langle {\cal O}_{\rm h} \rangle_{t=1} 
	\times \left[1+\sum_{x_0} t_1(x_0) g_0^2 \right]\,.
\ees
Here we have written down the expression for ${\cal O}_{\rm h}$ which
contains a single heavy quark, otherwise several terms $\propto t_1$
would appear. The function $t_1$ is the result of the 
sum over the loop momentum of the
tadpole graphs. Due to translation invariance in space it depends only
on the time coordinate $x_0$. The sum over $x_0$ in \eq{e:Ot}
extends over all timeslices
over which the heavy quark propagates. If, for the simplicity of the argument,
we assume that we also have translation invariance in time, $t_1$ does not
depend on $x_0$ and is a pure (dimensionless) number (note that we
do not have a factor $a$ in front of the sum).
It is then evident that $\eselfone$ is shifted by an amount $t_1$,
while $\delta V^{(1)}(0)$ is shifted by $2\,t_1$ and
$r^{(1)}$ is unchanged as claimed.

As (at one-loop order and with periodic boundary
conditions) the effect of $t$ can entirely be absorbed
into a change of $\eselfone$, it is also clear that the
improvement coefficients $\castat,\bastat$ are independent
of $t$ and in particular they  are identical for $S_{\rm h}^{\rm A}$
and $S_{\rm h}^{\rm s}$.

\subsection{Improvement coefficients}

To compute the improvement coefficients of the static-light axial current we
make a one loop perturbative expansion of a correlator involving the
static-light O($a$) improved axial current
\be
\fastatimpr(x_0) = \fastat(x_0) + a~\castat \fdeltaastat(x_0)\; .
\ee
The perturbative expansion reads
\be
\fastat(x_0) = \sum_{k=0}^{\infty}g_0^{2k}f_{\rm A}^{{\rm stat},(k)}(x_0)\; ,
\ee
and the Feynman diagrams contributing to the one loop term $f_{\rm A}^{{\rm
stat},(1)}$ are summarized in figure~B.1 of \cite{zastat:pap1}.

In the following, if not explicitly indicated,
we will insert the local operators always at $x_0=T/2$ and supress that argument.

\subsection{Determination of $\castat$}

The universality of the continuum limit 
gives us a handle to compute 
the improvement coefficient for a certain static action, once the improvement
coefficient is known for another static action.
This remark, given two actions with the same continuum
limit, is valid in general and not restricted to the static case.
It can be translated into the fact that the ratio 
\be
\rho_{\rm I}(S_{\rm h};\theta,\theta') = {\fastatimpr(S_h)|_{\theta} \over
\fastatimpr(S_h)|_{\theta'}}\; ,
\label{eq:rho}
\ee
is independent
of the action used up to O($a^2$), once suitable values of the 
improvement coefficients are chosen. 
The basic idea is then to use the already known value of 
$c_{\rm A}^{{\rm stat},(1)}$ 
for the EH action \cite{zastat:pap1,castat} and
to enforce
\be
\rho_{\rm I}(S_{\rm h}^{\rm EH};\theta,\theta') = 
\rho_{\rm I}(S_{\rm h}^{\rm A};\theta,\theta')\; ,
\label{eq:impr_cond2}
\ee
in order to compute $c_{\rm A}^{{\rm stat},(1)}$ for $S_{\rm h}^{\rm A}$.
Different choices for $\theta$ and $\theta'$ will give equivalent
definitions for the improvement coefficient, differing only by cutoff effects
$\rmO(a/L)$.
The perturbative expansion of eq. (\ref{eq:rho}) reads
\be
\rho_{\rm I}(S_{\rm h};\theta,\theta') =
\sum_{k=0}^{\infty}g_0^{2k}\rho_{\rm I}^{(k)}(S_{\rm h};\theta,\theta')\; .
\ee
After some trivial algebra we obtain
\be
\rho_{\rm I}^{(1)}(S_{\rm h};\theta,\theta') = \rho^{(1)}(S_{\rm
h};\theta,\theta') + ac_{\rm A}^{{\rm stat},(1)}(S_{\rm h})
\Big[\Delta(\theta) - \Delta(\theta') \Big]\; ,
\ee
and
\be
\rho^{(1)}(S_{\rm h};\theta,\theta') = {f_{\rm A}^{{\rm
stat},(1)}(S_{\rm h})|_{\theta} \over f_{\rm A}^{{\rm stat},(0)}|_{\theta}} - 
{f_{\rm A}^{{\rm stat},(1)}(S_{\rm h})|_{\theta'} \over f_{\rm A}^{{\rm stat},(0)}|_{\theta'}} 
\label{rho1}\; ,
\ee
where we use the fact that the correlator at tree-level does not depend on the static action,
and
\be 
\Delta(\theta) = {f_{\delta {\rm A}}^{{\rm stat},(0)}|_{\theta} \over f_{\rm A}^{{\rm
stat},(0)}|_{\theta}}\; .
\ee
An explicit formula for the desired 
improvement coefficient is
\bes
c_{\rm A}^{{\rm stat},(1)}(S_{\rm h}^{\rm A}) &=& c_{\rm A}^{{\rm stat},(1)}(S_{\rm h}^{\rm EH}) - \lim_{a/L \rightarrow 0} \Delta c  
\nonumber \\[-1ex] \label{castat12} \\[-1ex] \nonumber
 \Delta c  &=& 
{ \rho^{(1)}(S_{\rm h}^{\rm A};\theta,\theta') - \rho^{(1)}(S_{\rm h}^{\rm EH};\theta,\theta') \over
\Delta(\theta) - \Delta(\theta')} \; .
\ees
We have performed a one loop computation of $\Delta c$ 
for $\theta,\theta' \in \{0.0,0.5,1.0\}$,  $T=L$, and for resolutions
$L/a=4,6,\ldots,48$. The bare light
quark mass was set to the critical mass
\be
\mcrit=\mcritone g_0^2 + O(g_0^4) \; ,
\ee
with \cite{impr:pap2}
\be
a\mcritone = -0.2025565 \times \cf \; .
\ee
The input value 
\be
c_{\rm A}^{{\rm stat},(1)}(S_{\rm h}^{\rm EH}) = - {1 \over 4\pi} \times 1.0351(1)\; ,
\ee
was taken from \cite{castat}.
\begin{figure}[htb]
\hspace{0cm}
\vspace{-0.0cm}
\hspace{-0.0cm}
\begin{center}
\psfig{file=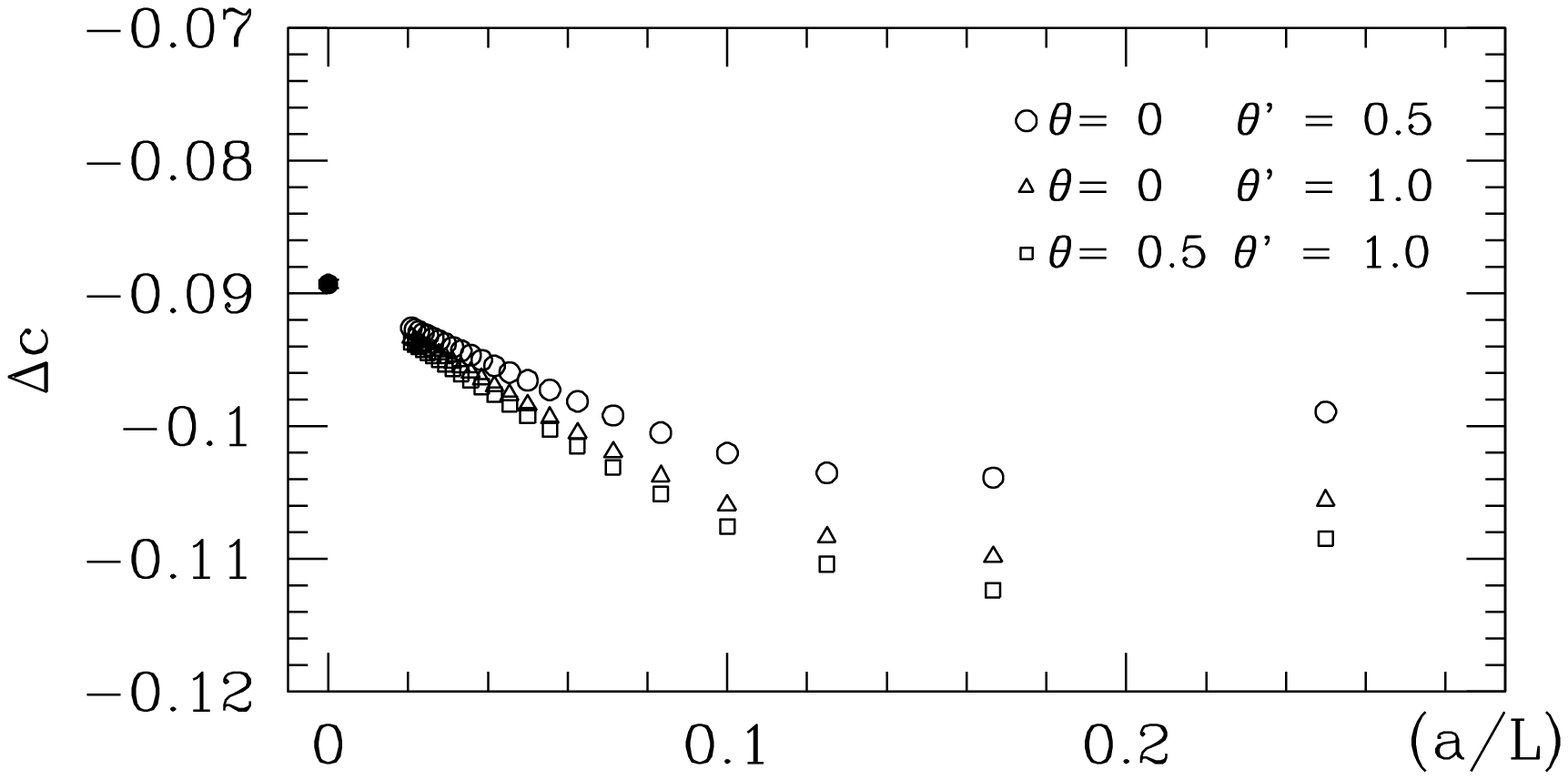,width=9cm,angle=0}
\end{center}
\vspace{-2.0cm}
\caption{\footnotesize\sl Dependence of 
   $\Delta c$, \protect\eq{castat12}, on $a/L$.}
\label{fig:cAeff}
\end{figure}
In figure~\ref{fig:cAeff} we plot $\Delta c$
as a function of $a/L$ for the $3$ possible choices of ($\theta$,$\theta'$).
It is clear that the limit $a/L \rightarrow 0$ can be taken with 
reasonable precision.
The plot indicates also that defining the improvement coefficient 
without performing the continuum limit  
will give a perturbative O($a/L$) uncertainty of the order of $0.02\,g_0^2$.
To obtain the final number \eq{caA} we used both the extrapolation procedures of
\cite{pt:LWeff} and \cite{Bucarelli}, finding consistent results.

\subsection{Determination of $\bastat$}

To determine $\bastat$ we follow the strategy just applied
to compute $c_{\rm A}^{{\rm stat}}$.
The ratio 
\be
\tilde\rho_{\rm I}(S_{\rm h};\mq',\mq) = 
{(1+ab_{\rm A}^{\rm stat}(S_{\rm h}) \mq')\fastatimpr(S_{\rm h})|_{\mq'} \over 
(1+ab_{\rm A}^{\rm stat}(S_{\rm h}) \mq)\fastatimpr(S_{\rm h})|_{\mq}}\; ,
\label{eq:rhotilde}
\ee
has a well defined continuum limit. Once suitable values of 
the improvement coefficients are chosen, 
the continuum limit is approached with a rate
of O($a^2$), as long as one considers (i)
the quenched approximation or (ii) is interested 
in one-loop accuracy, only. These restrictions are necessary,
since for full QCD and starting at order
$g_0^4$, additional terms proportional to
$a\mq$ with coefficients denoted by $b_{\zeta_{\rm h}},b_g$ (see \cite{zastat:pap1})
are needed to cancel all $\rmO(a)$ effects (the term proportional
to $b_\zeta$ cancels trivially in the above ratio).  

We can hence require 
\be
\tilde\rho_{\rm I}(S_{\rm h}^{\rm EH};\mq',\mq) = 
\tilde\rho_{\rm I}(S_{\rm h}^{\rm A};\mq',\mq) +\rmO(g_0^4)\,,
\label{eq:ba_cond}
\ee
which yields immediately
\bea
b_{\rm A}^{\rm stat,(1)}(S_{\rm h}^{\rm A}) &=& b_{\rm A}^{\rm stat,(1)}(S_{\rm h}^{\rm EH}) + \lim_{a/L\to0} \Delta b\,, \\
\label{e:deltab}
 \Delta b &=& 
{1 \over a(\mq'-\mq)}
\Big\{\Big[{f_{\rm A,I}^{\rm stat,(1)}(S_{\rm h}^{\rm EH})|_{\mq'} \over f_{\rm
A}^{\rm stat,(0)}(S_{\rm h}^{\rm EH})|_{\mq'}} - 
{f_{\rm A,I}^{\rm stat,(1)}(S_{\rm }h^{\rm EH})|_{\mq} \over f_{\rm A}^{\rm
    stat,(0)}(S_{\rm h}^{\rm EH})|_{\mq}}\Big] - \nonumber \\ 
&& \Big[{f_{\rm A,I}^{\rm stat,(1)}(S_{\rm h}^{\rm A})|_{\mq'} \over f_{\rm A}^{\rm
stat,(0)}(S_{\rm h}^{\rm A})|_{\mq'}} - 
{f_{\rm A,I}^{\rm stat,(1)}(S_{\rm h}^{\rm A})|_{\mq} \over f_{\rm A}^{\rm
    stat,(0)}(S_{\rm h}^{\rm A})|_{\mq}}\Big] \Big\}\; .
\label{bastat12}
\eea
In the numerical evaluation, we set $\mq=0$ and choose
$\mq'$ such that
\be
L\mr' = L(\zm \mq' (1+\bm a \mq')) = 0.24\; .
\label{cost_phys}
\ee
Here, in contrast to the MC evaluation, $\mr'$ has been defined 
in the lattice minimal subtraction scheme, where 
$\zm=1+\frac{1}{2\pi^2}\ln(L/a)g_0^2$.
The improvement constant $\bm=-\frac12 - 0.07217\, \cf\, g_0^2$
is known from \cite{pert:1loop} and
the input value of $b_{\rm A}^{\rm stat,(1)}$ for the EH action 
\be
b_{\rm A}^{\rm stat,(1)}(S_{\rm h}^{\rm EH}) = -0.056(7)\; ,
\ee
was computed in \cite{zastat:pap1}.

\begin{figure}[htb]
\hspace{0cm}
\vspace{-0.0cm}
\hspace{-0.0cm}
\begin{center}
\psfig{file=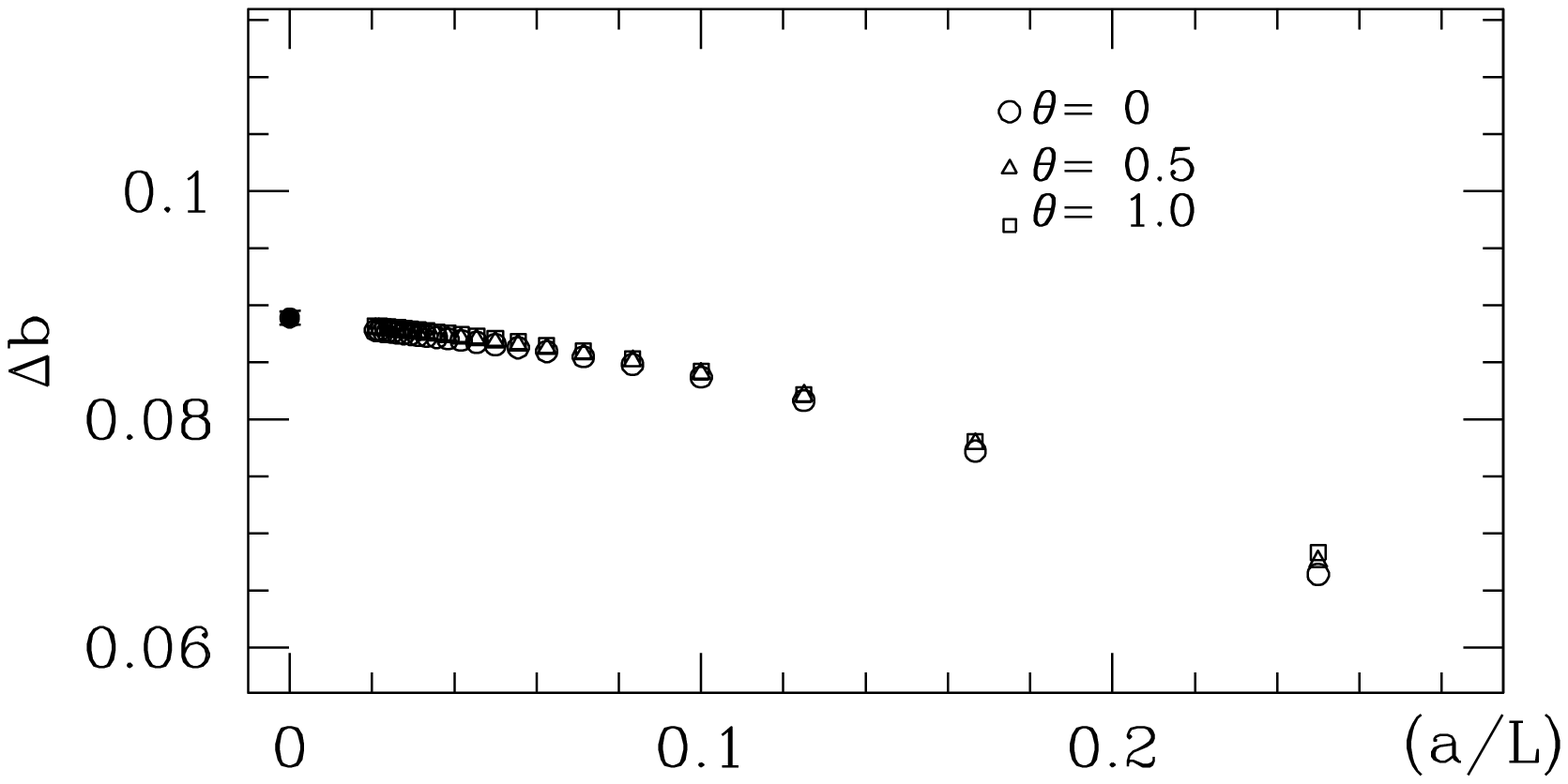,width=10cm,angle=0}
\end{center}
\vspace{-2.0cm}
\caption{\footnotesize\sl Dependence of $\Delta b$, \eq{e:deltab}, on $a/L$.}
\label{fig:bA_APE}
\end{figure}
In figure~\ref{fig:bA_APE} we plot $\Delta b$ 
as a function
of $a/L$ for the 3 possible choices of $\theta$. Again 
the O($a/L$) effects 
are well under control and almost independent of the choice of $\theta$. 
Extrapolating $a/L\to 0$ as above yields the result \eq{baA}.

\section{
Approximate one-link integral
\label{s:onelink}}

The basic idea of the one-link integral is that (in the pure
gauge theory) correlation functions of Polyakov loops remain unchanged if the
link variable $U(x,0)$ is replaced by
\be
\overline{U}(x,0) = {{\int {\rm d}U \; U\; \exp\left(12/g_0^2 \; \Re \tr 
[UV^{\dagger}(x,0)] \right)}\over  {\int {\rm d}U \exp\left(12/g_0^2 \; \Re \tr 
[UV^{\dagger}(x,0)] \right)}} \; ,
\ee
whereas the variance (and therefore the error) is substantially reduced.
While $\overline{U}(x,0)$ would thus be a possible choice for
$W(x,0)$, its expression is rather unhandy for gauge groups SU($N$) with $N>2$, 
\cite{integrals,deFo}. A more convenient approximation can be inferred form
the  SU($2$) case, where the group integral is proportional to
$V(x,0)$ and takes the form
\be
\overline{U}(x,0)={{V}\over{v}} {{I_2(24v/g_0^2)}\over{I_1(24v/g_0^2)}} \;,
\quad v=\left\{ {{1}\over{N}} \tr V V^{\dagger} \right\}^{1/2}\;, 
\quad V\equiv V(x,0) \;, 
\ee
in terms of the Bessel functions $I_1$ and $I_2$. For SU($3$) we
make a similar ansatz
\be
W(x,0)= g_0^{-2} V(x,0) s(g_0^{-2}v) \; .
\label{slink}
\ee
Although this can not reproduce the SU($3$) one-link integral exactly, since 
that is a more complicated function of $V$ and $V^{\dagger}$,
it turns out that the simple form
\be
s(y) =k/(y+1/5) \;,
\ee
with $k=1.092$ comes very close and thus is a good candidate to define an action
with reduced noise and roughly unchanged signal with respect to the Eichten-Hill
action.
The form of $s(y)$ has been  obtained numerically 
by minimizing  the difference between the norm 
of the links $\overline{U}(x,0)$ (constructed by a multihit procedure) 
and the norm of the links in \eq{slink} with $s(y)$ replaced by some
trial functions of $y$.
For this tuning we used  a set of quenched configurations
with SF boundary conditions on an $8^4$ lattice at $\beta=6$ and
$\beta=6.5$.
\footnote{The actual value of $k$ is irrelevant for physics as it only 
influences the self energy. We chose $k=1$ in \eq{Ss}.}

\section{Simulation parameters \label{s:param}}

We list in table~\ref{param} the lattice sizes and the bare parameters
of our simulations. The angle $\theta$ defines the periodicity of the
fermionic fields in all the spatial directions. We set $T/L=1$ and choose 
the background field to vanish.
\begin{table}[tb]
\centering
\begin{tabular}{c|ccccc}
\hline\\[-2.0ex]
run   &   $L/a$      &   $\beta$   &   $\kappa$   &    $\theta$ & $\#$ conf.\\[.5ex]
\hline\\[-2.ex]
 I    &    8         &  6.0219     & 0.135081,~0.1344011  &  0.0,~0.5,~1.0 & 4800 \\[.7ex]
 II   &    10        &  6.1628     & 0.135647,~0.1351239  &  0.0,~0.5,~1.0 & 3520 \\[.7ex]
 III  &    12        &  6.2885     & 0.135750,~0.1353237  &  0.0,~0.5,~1.0 & 4000 \\[.7ex]
 IV   &    16        &  6.4956     & 0.135593,~0.1352809  &  0.0,~0.5,~1.0 & 3200 \\[.7ex]
 ZI   &    6         &  6.2204     &     0.135470         &      0.5       & 4000 \\[.7ex]
 sZI  &    12        &  6.2204     &     0.135470         &      0.5       & 7200 \\[.7ex]
 ZII  &    8         &  6.4527     &     0.135543         &      0.5       & 3200 \\[.7ex]
 sZII &    16        &  6.4527     &     0.135543         &      0.5       & 3200 \\[.7ex]
 ZIII &    12        &  6.7750     &     0.135121         &      0.5       & 7360 \\[.7ex]
sZIII &    24        &  6.7750     &     0.135121         &      0.5       & 2000 \\[.7ex]
\hline 
\end{tabular}
\caption{\footnotesize{\sl Collection of the simulation parameters and statistics.}}
\label{param}
\end{table}
The first $\kappa$ value always corresponds to $\kappa_{\rm c}$, 
where the PCAC mass vanishes \cite{zastat:pap3,mbar:pap1}, while
 the second one for the runs I to IV has been chosen  
such that $Lm_{\rm R} =0.24$ with $m_{\rm R}$ the running
quark mass in the SF scheme at the scale $\mu^{-1} = 2 L_{\rm max}
=1.436 r_0$:
\begin{equation}
m_{\rm R} = Z_m \widetilde{m}_{\rm q}\;, \quad \widetilde{m}_{\rm q}=m_{\rm q}(1+
b_{\rm m} am_{\rm q})\; , \quad am_{\rm q}={{1}\over{2}} \left(
{{1}\over{\kappa}} -{{1}\over{\kappa_{\rm c}}} \right) \; .
\label{qmass}
\end{equation}
The low energy scale $r_0$ ($\simeq 0.5$ fm) has been introduced in 
\cite{pot:r0} and in \cite{pot:r0_SU3} the ratio $r_0/a$ has been computed 
to a precision of about $0.5 \%$ in a range of  $\beta$ values 
which includes runs I to IV in table~\ref{param}.  
For the coefficients  $Z_m$ and $b_{\rm m}$ in~\eq{qmass} we have used 
the results in~\cite{impr:babp} and the improvement coefficients
$\ct,\cttilde$ were set to their two-loop and one-loop values 
respectively~\cite{SF:LNWW,pert:2loop_nf0,impr:pap5}.

\end{appendix}


\begin{thebibliography}{10}

\bibitem{CKM:CERN}
M. Battaglia et~al.,
\newblock (2003), hep-ph/0304132.

\bibitem{algo:GHMC}
M. Hasenbusch,
\newblock Phys. Lett. B519 (2001) 177.

\bibitem{algo:GHMC3}
M. Hasenbusch and K. Jansen,
\newblock Nucl. Phys. B659 (2003) 299, hep-lat/0211042.

\bibitem{algo:L1}
M. L{\"u}scher,
\newblock JHEP 05 (2003) 052, hep-lat/0304007.

\bibitem{algo:L2}
M. {L\"u}scher,
\newblock Comput. Phys. Commun. 165 (2005) 199, hep-lat/0409106.

\bibitem{lat91:lepage}
G.P. Lepage,
\newblock Nucl. Phys. Proc. Suppl. 26 (1992) 45.

\bibitem{lat03:kronfeld}
A.S. Kronfeld,
\newblock Nucl. Phys. Proc. Suppl. 129 (2004) 46, hep-lat/0310063.

\bibitem{reviews:physcoll}
R. Sommer,
\newblock ECONF C030626 (2003) FRAT06, hep-ph/0309320.

\bibitem{hqet:pap1}
ALPHA, J. Heitger and R. Sommer,
\newblock JHEP 02 (2004) 022, hep-lat/0310035.

\bibitem{lat03:juri}
ALPHA, J. Rolf et~al.,
\newblock Nucl. Phys. Proc. Suppl. 129 (2004) 322, hep-lat/0309072.

\bibitem{stat:eichten}
E. Eichten,
\newblock Talk delivered at the Int. Sympos. of Field Theory on the Lattice,
  Seillac, France, Sep 28 - Oct 2, 1987.

\bibitem{stat:hashi}
S. Hashimoto,
\newblock Phys. Rev. D50 (1994) 4639, hep-lat/9403028.

\bibitem{fbstat:old1}
C.R. Allton, C.T. Sachrajda, V. Lubicz, L. Maiani and G. Martinelli,
\newblock Nucl. Phys. B349 (1991) 598.

\bibitem{fbstat:old2}
C. Alexandrou, F. Jegerlehner, S. {G\"usken}, K. Schilling and R. Sommer,
\newblock Phys. Lett. B256 (1991) 60.

\bibitem{stat:fnal2}
A. Duncan et~al.,
\newblock Phys. Rev. D51 (1995) 5101, hep-lat/9407025.

\bibitem{stat:eichhill1}
E. Eichten and B. Hill,
\newblock Phys. Lett. B234 (1990) 511.

\bibitem{reviews:beauty}
R. Sommer,
\newblock Phys. Rept. 275 (1996) 1, hep-lat/9401037.

\bibitem{stat:letter}
ALPHA, M. Della~Morte et~al.,
\newblock Phys. Lett. B581 (2004) 93, hep-lat/0307021.

\bibitem{SF:LNWW}
M. {L\"uscher}, R. Narayanan, P. Weisz and U. Wolff,
\newblock Nucl. Phys. B384 (1992) 168, hep-lat/9207009.

\bibitem{SF:stefan1}
S. Sint,
\newblock Nucl. Phys. B421 (1994) 135, hep-lat/9312079.

\bibitem{zastat:pap1}
ALPHA, M. Kurth and R. Sommer,
\newblock Nucl. Phys. B597 (2001) 488, hep-lat/0007002.

\bibitem{impr:SW}
B. Sheikholeslami and R. Wohlert,
\newblock Nucl. Phys. B259 (1985) 572.

\bibitem{impr:pap1}
M. {L\"uscher}, S. Sint, R. Sommer and P. Weisz,
\newblock Nucl. Phys. B478 (1996) 365, hep-lat/9605038.

\bibitem{impr:pap3}
M. {L\"uscher}, S. Sint, R. Sommer, P. Weisz and U. Wolff,
\newblock Nucl. Phys. B491 (1997) 323, hep-lat/9609035.

\bibitem{lat03:michele}
ALPHA, M. Della~Morte et~al.,
\newblock Nucl. Phys. Proc. Suppl. 129 (2004) 346, hep-lat/0309080.

\bibitem{stat:eichhill_za}
E. Eichten and B. Hill,
\newblock Phys. Lett. B240 (1990) 193.

\bibitem{PPR}
G. Parisi, R. Petronzio and F. Rapuano,
\newblock Phys. Lett. 128B (1983) 418.

\bibitem{hyp}
A. Hasenfratz and F. Knechtli,
\newblock Phys. Rev. D64 (2001) 034504, hep-lat/0103029.

\bibitem{HYP:pot}
A. Hasenfratz, R. Hoffmann and F. Knechtli,
\newblock Nucl. Phys. Proc. Suppl. 106 (2002) 418, hep-lat/0110168.

\bibitem{smear:ape}
APE, M. Albanese et~al.,
\newblock Phys. Lett. 192B (1987) 163.

\bibitem{impr:Sym1}
K. Symanzik,
\newblock Nucl. Phys. B226 (1983) 187.

\bibitem{impr:Sym2}
K. Symanzik,
\newblock Nucl. Phys. B226 (1983) 205.

\bibitem{MORNINGSTAR1}
C. Morningstar and J. Shigemitsu,
\newblock Phys. Rev. D57 (1998), hep-lat/9712015.

\bibitem{castat}
K.I. Ishikawa, T. Onogi and N. Yamada,
\newblock Nucl. Phys. Proc. Suppl. 83 (2000) 301, hep-lat/9909159.

\bibitem{zastat:pap3}
ALPHA, J. Heitger, M. Kurth and R. Sommer,
\newblock Nucl. Phys. B669 (2003) 173, hep-lat/0302019.

\bibitem{lat01:rainer}
ALPHA, J. Heitger and R. Sommer,
\newblock Nucl. Phys. Proc. Suppl. 106 (2002) 358, hep-lat/0110016.

\bibitem{pot:r0}
R. Sommer,
\newblock Nucl. Phys. B411 (1994) 839, hep-lat/9310022.

\bibitem{pot:r0_SU3}
ALPHA, M. Guagnelli, R. Sommer and H. Wittig,
\newblock Nucl. Phys. B535 (1998) 389, hep-lat/9806005.

\bibitem{impr:babp}
ALPHA, M. Guagnelli et~al.,
\newblock Nucl. Phys. B595 (2001) 44, hep-lat/0009021.

\bibitem{impr:pap5}
S. Sint and P. Weisz,
\newblock Nucl. Phys. B502 (1997) 251, hep-lat/9704001.

\bibitem{alpha:SU3}
M. {L\"uscher}, R. Sommer, P. Weisz and U. Wolff,
\newblock Nucl. Phys. B413 (1994) 481, hep-lat/9309005.

\bibitem{mbar:pap1}
ALPHA, S. Capitani, M. {L\"uscher}, R. Sommer and H. Wittig,
\newblock Nucl. Phys. B544 (1999) 669, hep-lat/9810063.

\bibitem{pot:intermed}
S. Necco and R. Sommer,
\newblock Nucl. Phys. B622 (2002) 328, hep-lat/0108008.

\bibitem{algo:multilevel}
M. Luscher and P. Weisz,
\newblock JHEP 09 (2001) 010, hep-lat/0108014.

\bibitem{pot:sesam}
SESAM, G.S. Bali, H. Neff, T. Duessel, T. Lippert and K. Schilling,
\newblock (2005), hep-lat/0505012.

\bibitem{lat04:stephan}
S. {D\"urr}, A. {J\"uttner}, J. Rolf and R. Sommer,
\newblock (2004), hep-lat/0409058.

\bibitem{impr:pap2}
M. {L\"uscher} and P. Weisz,
\newblock Nucl. Phys. B479 (1996) 429, hep-lat/9606016.

\bibitem{pt:LWeff}
M. L{\"u}scher and P. Weisz,
\newblock Nucl. Phys. B266 (1986) 309.

\bibitem{Bucarelli}
A. Bucarelli, F. Palombi, R. Petronzio and A. Shindler,
\newblock Nucl. Phys. B552 (1999) 379, hep-lat/9808005.

\bibitem{pert:1loop}
S. Sint and R. Sommer,
\newblock Nucl. Phys. B465 (1996) 71, hep-lat/9508012.

\bibitem{integrals}
R. Brower, P. Rossi and C.I. Tan,
\newblock Nucl. Phys. B190 (1981) 699.

\bibitem{deFo}
P. De~Forcrand and C. Roiesnel,
\newblock Phys. Lett. B151 (1985) 77.

\bibitem{pert:2loop_nf0}
ALPHA, A. Bode, U. Wolff and P. Weisz,
\newblock Nucl. Phys. B540 (1999) 491, hep-lat/9809175.

\end{thebibliography}

\end{document}